\documentclass[10pt,twocolumn,english,aps,nofootnoteinbib,manuscript]{revtex4}
\usepackage[T1]{fontenc}
\usepackage[latin9]{inputenc}
\usepackage{color}
\usepackage{amsmath}
\usepackage{amssymb}
\usepackage{graphicx}
\usepackage{esint}

\makeatletter
\@ifundefined{textcolor}{}
{%
 \definecolor{BLACK}{gray}{0}
 \definecolor{WHITE}{gray}{1}
 \definecolor{RED}{rgb}{1,0,0}
 \definecolor{GREEN}{rgb}{0,1,0}
 \definecolor{BLUE}{rgb}{0,0,1}
 \definecolor{CYAN}{cmyk}{1,0,0,0}
 \definecolor{MAGENTA}{cmyk}{0,1,0,0}
 \definecolor{YELLOW}{cmyk}{0,0,1,0}
 }

\@ifundefined{definecolor}
 {\usepackage{color}}{}
\@ifundefined{definecolor}{\usepackage{color}}{}
\@ifundefined{definecolor}{\usepackage{color}}{}
\@ifundefined{definecolor}{\@ifundefined{definecolor}
 {\usepackage{color}}{}
}{}
\@ifundefined{definecolor}{\@ifundefined{definecolor}
 {\@ifundefined{definecolor}
 {\usepackage{color}}{}
}{}
}{}

\usepackage{babel}

\usepackage{babel}

\usepackage{babel}

\usepackage{babel}

\makeatother

\usepackage{babel}
\begin{document}

\title{Chiral filtering in graphene with coupled valleys}

\author{P. L. e S. Lopes$^{1}$, A. H. Castro Neto$^{2}$%
\thanks{On leave from Department of Physics, Boston University%
} and A. O. Caldeira$^{1}$ }

\affiliation{$^{1}$Instituto de Física Gleb Wataghin, Universidade Estadual de
Campinas, Campinas, SP 13083-970, Brazil}

\affiliation{$^{2}$Graphene Research Centre and Physics Department, National
University of Singapore, 2 Science Drive 3, Singapore, 117542}
\begin{abstract}
We analyze the problem of electronic transmission through semi-infinite
and finite regions of graphene which are characterized by different
types of connections between the Dirac points. These valley symmetry
breaking Hamiltonians might arise from electronic self-interaction
mediated by the dielectric environment of distinct parts of the substrate
on which the graphene sheet is placed. We show that it is possible
to have situations in which we can use these regions to select or
filter states of one desired chirality.
\end{abstract}
\maketitle

\section{Introduction}

Graphene is a truly two-dimensional (2D) crystal with Dirac-like quasi-particles
\cite{key-1,key-1b}. These charge carriers are the result of the
wavefunction interference due to the honeycomb lattice structure and
hence they cannot exist outside the many-body system (electrons-plus-lattice).
From this perspective, the low energy, long wavelength, Lorentz invariance
in this system is an emergent phenomenon and, as such, can be influenced
by external factors such as disorder, applied electric and magnetic
fields, and structural deformations of the lattice such as pressure,
strain, and shear \cite{key-2}.

The low energy physics of these Dirac quasi-particles is strongly
influenced by the fact that the two Dirac cones sit at the corners
of the hexagonal Brillouin zone (K and K'). These cones are related
by time reversal symmetry and define the \textit{chirality} of these
quasi-particles in terms of their momentum relative to each cone.
Unlike the case of neutrinos, where only one chiral flavor exists,
Dirac particles in graphene have both flavors, and they can be either
right-handed or left-handed depending on whether they reside in one
cone or the other. While in a perfect, non-interacting, graphene sheet
these chiral states are decoupled, in a real graphene sample they
can be coupled by the external perturbations mentioned above. At low
energies, one can have either \textit{intra-valley} processes, with
small momentum transfer, that preserve chirality, or \textit{inter-valley}
processes, with large momentum transfer, that mixes chiralities. It
is well-known that, in the presence of weak disorder, intra-valley
processes lead to weak \textit{anti-localization} effects \cite{key-3},
whereas inter-valley processes lead to weak \textit{localization}
\cite{key-4}. Hence, the coupling between K and K' points play an
important role in the physics of graphene.

In this work we study the transport through regions where the K and
K' points are not coupled and regions in which they are actually coupled.
Although the Hamiltonians with which we deal are manifestly non interacting,
these couplings might arise from the interactions between the electrons,
as we argue below. Our main goal is to understand how the Dirac electrons
behave across the interfaces separating those two regions as a way
to classify the possible scattering mechanisms in graphene. Such a
situation can also be artificially created by depositing graphene
across substrates with different dielectric constants. For instance,
there are quantum Monte Carlo \cite{key-6} calculations that indicate
that suspended graphene (i.e., on vacuum with dielectric constant
$\epsilon_{0}=1$) is a excitonic insulator and graphene on SiO$_{2}$
($\epsilon_{{\rm SiO2}}\approx3$) should be a semi-metal \cite{key-7}.
An interface between those regions would have a transistor-like effect
with a large on-off ratio for current flow.

Our starting point is the well-known low energy effective Hamiltonian
for neutral graphene that is given by \cite{key-1} (we use units
such that $v_{F}=\hbar=1$):
\begin{equation}
\mathcal{H}=\int d^{2}r\left(\begin{array}{cc}
\hat{\Psi}_{1}^{\dagger} & \hat{\Psi}_{2}^{\dagger}\end{array}\right)\left(\begin{array}{cc}
\mathbf{\sigma}\cdot\mathbf{\hat{P}} & 0\\
0 & \mathbf{\sigma^{*}}\cdot\mathbf{\hat{P}}
\end{array}\right)\left(\begin{array}{c}
\hat{\Psi}_{1}\\
\hat{\Psi}_{2}
\end{array}\right),\label{eq:H2livre}
\end{equation}
 where $\hat{\Psi}_{i}^{\dagger}=\left(a_{i}^{\dagger},b_{i}^{\dagger}\right)$,
$a_{i}^{\dagger}$ and $b_{i}^{\dagger}$ being the creation operators
for electrons in sublattice A or B, respectively, in the Dirac cone
$i$ $\left(i=1,2\right)$ and $\mathbf{\sigma}=(\sigma_{x},\sigma_{y})$
is a 2D vector whose components are Pauli matrices (we ignore spin
variables).

In the representation of the Hamiltonian (\ref{eq:H2livre}) there
are many ways to couple the K and K' points \cite{key-7}. These couplings
have different symmetries and hence represent different physical processes.
Notice, however, that all the processes that couple the two cones
can be represented in terms of combinations of the identity matrix
and Pauli matrices in the "valley" space. We will deal only with
real coupling potentials in such a way that we can write explicitly
all the generators of the different couplings as combinations of basic
matrices:
\begin{equation}
\mathcal{V}_{x}=\Delta\left(\begin{array}{cc}
0 & \sigma_{x}\\
\sigma_{x} & 0
\end{array}\right)\,,\label{eq:v1}
\end{equation}

\begin{equation}
\mathcal{V}_{y}=\Delta\left(\begin{array}{cc}
0 & -i\sigma_{y}\\
i\sigma_{y} & 0
\end{array}\right)\,,\label{eq:v2}
\end{equation}

\begin{equation}
\mathcal{V}_{z}=\Delta\left(\begin{array}{cc}
0 & \sigma_{z}\\
\sigma_{z} & 0
\end{array}\right)\,,\label{eq:v3}
\end{equation}
 and
\begin{equation}
\mathcal{V}_{I}=\Delta\left(\begin{array}{cc}
0 & 1\\
1 & 0
\end{array}\right)\,,\label{eq:v4}
\end{equation}
 where $\Delta$ is a parameter that is assumed to be positive throughout
the paper since negative deltas will bring no new physics. Of course
the most arbitrary potential would be a linear combination of these
with different delta parameters (and in this case the sign difference
between them might be relevant).

As we are going to show, only $\mathcal{V}_{y}$ and $\mathcal{V}_{x}$
are of real interest and will be the focus of our studies. We will
also show that the transmission through regions described by these
matrices presents very unusual properties and can presumably be measured
experimentally (we stress, nevertheless, that it is not our aim to fully
solve the transport problem considering
finite size and  disorder effects, but to analyze the 
physics of our proposed model which we believe can account for the main properties
behind, for instance, the contact resistance of nanoscopic graphene junctions).

The paper is organized as follows. In Sec. II we give a very
brief introduction to the problem of "free" graphene and some of
our notation. In Sec. III we present and analyze the Hamiltonians
with which we work, giving a somewhat general notion on why our
choices of interactions and their nature, solving the
corresponding eigenvalue problems and analyzing the symmetries
they obey. In Sec. IV, we (less) briefly discuss the transmission
problem in two dimensions. In Sec. V we solve the problem of
transmission through the interface between two semi-infinite
regions of graphene, one with "free" electrons and the other with
self-interacting quasi-particles and finally, in Sec. VI, we solve
the finite-barrier-like problem of transmission through a finite
region of the interacting material. In Sec. VII, we try to make a
connection of our results and the Landauer formalism. In Sec. VIII
we present our conclusions.

\section{Basic Properties}

The eigenvalue problem given by (\ref{eq:H2livre}) can be solved
by going into the momentum representation
\begin{equation}
\left(\begin{array}{cc}
\mathbf{\sigma}\cdot\mathbf{k} & 0\\
0 & \mathbf{\sigma}^{*}\cdot\mathbf{k}
\end{array}\right)\psi_{\pm\mathbf{K}(\mathbf{K'})}(\mathbf{k})=E_{\pm}\psi_{\pm\mathbf{K}(\mathbf{K'})}(\mathbf{k})
\end{equation}
 with energy eigenvalues, $E=\pm k$, and eigenstates of well defined
momenta around the Dirac points (with their corresponding spinors,
see below). Going back to the position representation one has $\psi_{\mathbf{K}(\mathbf{K'})}(\mathbf{r})=\psi_{\pm\mathbf{K}(\mathbf{K'})}(\mathbf{k})e^{i\mathbf{k}\cdot\mathbf{r}}$
where the spinors are
\begin{eqnarray}
\psi_{\pm\mathbf{K}}(\mathbf{k}) & = & \frac{1}{\sqrt{2}}\left(\begin{array}{c}
e^{-i\theta/2}\\
\pm e^{i\theta/2}\\
0\\
0
\end{array}\right),\nonumber \\
\psi_{\pm\mathbf{K'}}(\mathbf{k}) & = & \frac{1}{\sqrt{2}}\left(\begin{array}{c}
0\\
0\\
e^{i\theta/2}\\
\pm e^{-i\theta/2}
\end{array}\right),\label{eq:mudabase}
\end{eqnarray}
 with $\theta=\arctan(k_{y}/k_{x})$ and the vector $\mathbf{k}$,
whose modulus is connected to the Fermi energy by the dispersion relation,
is centered either around $\mathbf{K}$ or $\mathbf{K'}$, respectively.
These eigenstates, which are symmetric and anti-symmetric linear combinations
of the states referring to the A and B sublattices, allow us to introduce
a particle-hole representation which reads
\begin{equation}
\mathcal{H}=\sum_{i=1,2}k\left[f_{k,i}^{\dagger}f_{k,i}-h_{k,i}^{\dagger}h_{k,i}\right],
\end{equation}
 where the hole states have negative energy. Thus, we describe the
electron behavior in the free graphene by a theory of non-interacting
massless fermionic quasi-particles with two different "flavors"
(and their corresponding anti-particles).

\section{Interacting Hamiltonians}

Electron-electron interaction plays an important role in graphene
physics and transport problems \cite{key-8a,key-8b}. Its effects
depend strongly on the dielectric function and is independent of
the electronic density \cite{key-8a}. Since the dielectric
function can be affected in many ways, as, for instance, by
adsorbed atoms on the surface or deposition substrates, we propose
that there may be situations in which the resulting effective
electron-electron interaction can cause mixing of the Dirac cones
and we show how it could happen, at least in a heuristic
mean-field approximation. Our approach is phenomenological,
and, therefore, the specific shape of the interaction potentials
will not be given and should be chosen from microscopic arguments.
There are many possible situations, and one of them is the
possibility that the (Coulomb) interaction becomes screened
\cite{key-8a,key-8c}. Hence, we justify the choice of our approach
and will make the symmetry assumptions we find reasonable to
achieve our results.

So, let us suppose that the  electron-electron interaction
contribution to the usual tight binding Hamiltonian is of the form
\begin{equation}
H_{I}=\sum_{i\neq j}U_{ij}n_{i}n_{j}.
\end{equation}
Separating the terms in the summation corresponding to each
sublattice, we arrive at
\begin{align}
H_{IA} & =\sum_{i\neq j}U_{ij}n_{Ai}n_{Aj}\\
H_{IB} & =\sum_{i\neq j}U_{ij}n_{Bi}n_{Bj}\\
H_{IAB} & =\sum_{i,j}V_{ij}n_{Ai}n_{Bj}.
\end{align}
We have assumed that the interaction between the electrons within
each sublattice is equal, which actually need not be.

Now we perform the  mean field approximation. The values of the
mean-fields should actually be determined in a self-consistent
way, which will not be done here, and be justified by microscopic
arguments. Our  point is that if there is any physical process
which gives rise to these mean-fields, the proposed decoupling
will lead to our results. We make more comments on the physical
nature of this approximation further below. That being said, we
can choose, for example, the first Hamiltonian and get, to first
order in the deviations from a mean density of electrons in
sublattice A,
\begin{align}
H_{IA} & \approx\sum_{i\neq j}U_{ij}\left[n_{Ai}\left\langle n_{Aj}\right\rangle +n_{Aj}\left\langle n_{Ai}\right\rangle -\left\langle n_{Ai}\right\rangle \left\langle n_{Aj}\right\rangle \right]\nonumber \\
 & =\sum_{i}\Lambda_{i}n_{Ai}-const.,
\end{align}
where $\Lambda$ is the effective on-site potential generated by
the self-interaction between the particles. In Fourier space, we
can write
\begin{equation}
a_{i}=\frac{1}{\sqrt{N_{S}}}\sum_{\mathbf{k}}e^{i\mathbf{k}\cdot\mathbf{R}_{i}}a_{\mathbf{k}},
\end{equation}
where $N_{S}$ is the number of sites on the lattice. Hence, we have,
\begin{equation}
H_{IA}\approx\sum_{\mathbf{k}\mathbf{k}'}\Delta_{\mathbf{k}-\mathbf{k}'}a_{\mathbf{k}}^{\dagger}a_{\mathbf{k}'},
\end{equation}
where we have ignored the constant term. Therefore, this
Hamiltonian couples the $\mathbf{k}$ vectors . If we keep only the the long wavelength contribution
to the expression just above, we find that this term leads to
the four couplings (following our notation,
$a_{\mathbf{K}}\rightarrow a_{1}$,$a_{\mathbf{K}'}\rightarrow
a_{2}$)
\begin{equation}
a_{1}^{\dagger}a_{1};\ a_{2}^{\dagger}a_{2}
\end{equation}
and
\begin{equation}
a_{1}^{\dagger}a_{2};\ a_{2}^{\dagger}a_{1}.
\end{equation}
The first two lead to diagonal terms which are of no
interest to us. The two other give contributions exactly of the type
we want. The same arguments can be carried over to $H_{IB}$.
Since the  potential must be such that $U_{ij}=U_{ji}$,  its Fourier transform is real. The mean fields should
also be real and this justifies our choice of only analyzing the four real
$\mathcal{V}$ Hamiltonians. As for the physical realization of these mean
fields, we believe that the presence of the substrate might, in specific
cases, affect the electrons in the
material and break the symmetries between the sub-lattices in such a
way that ordered states like, for example, a CDW could develop. More thought in this direction
is needed but will not be pursued in this paper.

If one wants a more thorough approach, one can use the Alicea
and Fisher model \cite{key-8d}. There it is shown how the long
and short range interactions give rise to terms like staggered densities
in each sublattice,
\begin{equation}
\rho_{stag}=n_{A1}+n_{A2}-n_{B1}-n_{B2},
\end{equation}
where $n_{A,Bi}$, $i=1,2$, refers to the number operator in the
sublattice A or B and Dirac cone 1 or 2. In the Alicea and Fisher
model, the long wavelength Hamiltonian has (local) contributions due to
electronic densities given by
\begin{eqnarray}
\rho_{tot}^{2}+\rho_{stag}^{2} & = & a_{1}^{\dagger}a_{1}a_{1}^{\dagger}a_{1}+a_{2}^{\dagger}a_{2}a_{2}^{\dagger}a_{2}+\nonumber \\
 &  & +b_{1}^{\dagger}b_{1}b_{1}^{\dagger}b_{1}+b_{2}^{\dagger}b_{2}b_{2}^{\dagger}b_{2}+b_{1}^{\dagger}b_{1}b_{2}^{\dagger}b_{2}+\nonumber \\
 &  & +a_{2}^{\dagger}a_{2}a_{1}^{\dagger}a_{1}+a_{1}^{\dagger}a_{1}a_{2}^{\dagger}a_{2}+b_{2}^{\dagger}b_{2}b_{1}^{\dagger}b_{1}.\nonumber\\
\end{eqnarray}
This term comes from the on-site repulsion interaction (a more detailed
analisys can be found in the original paper \cite{key-8d}). We can see that
a mean-field decoupling might also generate our Hamiltonians in this case,
however, the mean-fields in this case result directly from the replacement $a_{1}a_{2}^{\dagger}\rightarrow\left\langle a_{1}a_{2}^{\dagger}\right\rangle $
, for example. We find this type of mean-field harder to justify physically but
leave it here as an example.

As we had formerly proposed, we have given some heuristic arguments for our \textit{ad hoc} choice of the form of the $\mathcal{V}_i $ interactions.
We proceed now to exploit the  consequences of these model Hamiltonians in some specific cases.

\subsubsection*{Spectrum in the presence of $\mathcal{V}_{y}$ }

In the presence of the disturbance \eqref{eq:v2} the Hamiltonian
is given by
\begin{equation}
H=\left(\begin{array}{cccc}
0 & ke^{-i\theta} & 0 & -\Delta\\
ke^{i\theta} & 0 & \Delta & 0\\
0 & \Delta & 0 & ke^{i\theta}\\
-\Delta & 0 & ke^{-i\theta} & 0
\end{array}\right).\label{eq:AB'(-)BA'(+)}
\end{equation}

\begin{figure}[t]

\begin{centering}
\includegraphics[scale=0.7]{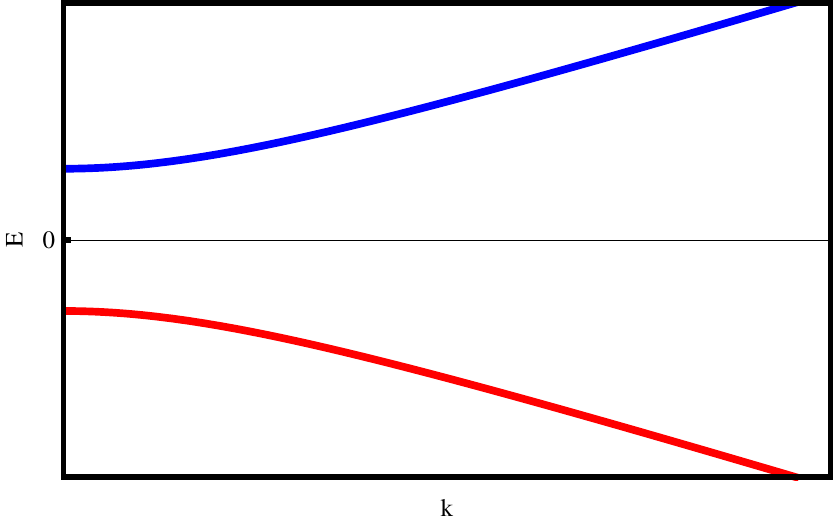}
\par\end{centering}

\caption{(color on-line) Schematics of the dispersion relation for the Hamiltonian
in the presence of the $V_{y}$ term (arbitrary units).}
\end{figure}

The solution of the eigenvalue problem gives $E_{A,B\pm}=\pm\sqrt{k^{2}+\Delta^{2}}\equiv\pm\epsilon$
(Fig. 1) with a gap of size $\Delta$ in the spectrum. The spinorial
part of the eigenvectors (we omit the plane waves for simplicity in
this whole section)
\begin{equation}
\left|\chi_{A}^{+}\right\rangle =\frac{1}{\sqrt{2}}\left(\begin{array}{c}
A_{+}e^{-i\theta/2}\\
A_{+}e^{i\theta/2}\\
A_{-}e^{i\theta/2}\\
-A_{-}e^{-i\theta/2}
\end{array}\right),\left|\chi_{A}^{-}\right\rangle =\frac{1}{\sqrt{2}}\left(\begin{array}{c}
A_{-}e^{-i\theta/2}\\
A_{-}e^{i\theta/2}\\
-A_{+}e^{i\theta/2}\\
A_{+}e^{-i\theta/2}
\end{array}\right),
\end{equation}

\begin{equation}
\left|\chi_{B}^{+}\right\rangle =\frac{1}{\sqrt{2}}\left(\begin{array}{c}
A_{-}e^{-i\theta/2}\\
-A_{-}e^{i\theta/2}\\
-A_{+}e^{i\theta/2}\\
-A_{+}e^{-i\theta/2}
\end{array}\right),\left|\chi_{B}^{-}\right\rangle =\frac{1}{\sqrt{2}}\left(\begin{array}{c}
A_{+}e^{-i\theta/2}\\
-A_{+}e^{i\theta/2}\\
A_{-}e^{i\theta/2}\\
A_{-}e^{-i\theta/2}
\end{array}\right),
\end{equation}
 where $A_{\pm}=\sqrt{\frac{\epsilon\pm k}{2\epsilon}}$. Here, $A$
and $B$ do not refer to the different triangular sub-lattices but
to the different degenerate states originating from the dispersion
relation.

as we saw, can use eq.\eqref{eq:mudabase} to put this Hamiltonian in a particle-hole
representation, where  particle and hole states are described as symmetric
and anti-symmetric combinations of the different sub-lattices wave
functions. The effective interaction between the particles
and holes is then written as
\begin{equation}
H_{ph}=\left(\begin{array}{cccc}
k & 0 & 0 & \Delta\\
0 & -k & -\Delta & 0\\
0 & -\Delta & k & 0\\
\Delta & 0 & 0 & -k
\end{array}\right),
\end{equation}
 where $ph$ refers to the particle-hole representation. Eq. \eqref{eq:AB'(-)BA'(+)}
can be seen to obey orthogonal time reversal symmetry (exchanging
the valleys by time reversal) and it is seen to generate a band gap. We see that this kind
of effective potential is equivalent to the introduction of an asymmetric
coupling between particles and holes from different Dirac cones.

\subsubsection*{Spectrum in the presence of $\mathcal{V}_{x}$}

In the presence of a perturbation of the form  (\ref{eq:v1}), the full
Hamiltonian becomes :
\begin{equation}
H=\left(\begin{array}{cccc}
0 & ke^{-i\theta} & 0 & \Delta\\
ke^{i\theta} & 0 & \Delta & 0\\
0 & \Delta & 0 & ke^{i\theta}\\
\Delta & 0 & ke^{-i\theta} & 0
\end{array}\right).\label{eq:AB'(+)BA'(+)}
\end{equation}

\begin{figure}[t]
\begin{centering}
\includegraphics[scale=0.7]{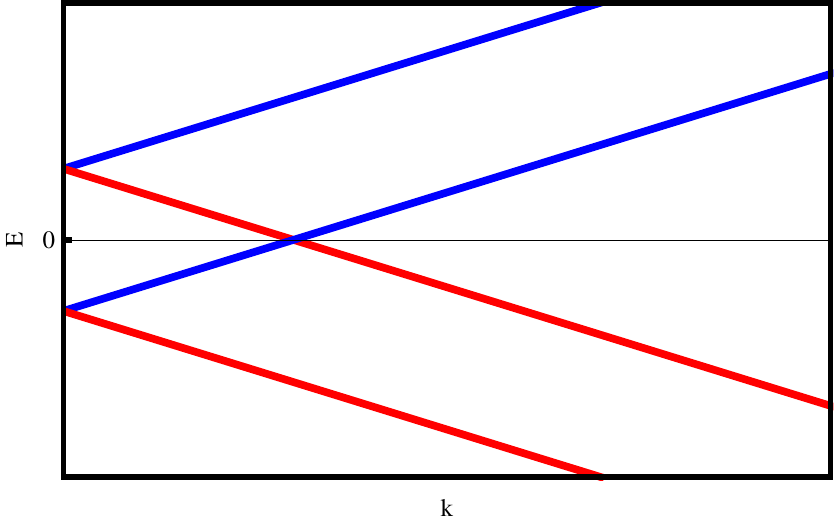}
\par\end{centering}

\centering{}

\caption{(color on-line) Schematics of the dispersion relation for the Hamiltonian
in the presence of the $V_{x}$ term(arbitrary units).}
\end{figure}

Diagonalization will lead us to $E_{\pm\pm}=\pm k\pm\Delta$ (Fig.
2) with no gap in the spectrum but a shift of the Dirac cones relative
to each other by $\Delta$. The eigenvectors are now
\begin{equation}
\left|\chi_{++}\right\rangle =\frac{1}{2}\left(\begin{array}{c}
e^{-i\theta/2}\\
e^{i\theta/2}\\
e^{i\theta/2}\\
e^{-i\theta/2}
\end{array}\right),\left|\chi_{+-}\right\rangle =\frac{1}{2}\left(\begin{array}{c}
e^{-i\theta/2}\\
e^{i\theta/2}\\
-e^{i\theta/2}\\
-e^{-i\theta/2}
\end{array}\right);\label{eq:partics int}
\end{equation}

\begin{equation}
\left|\chi_{-+}\right\rangle =\frac{1}{2}\left(\begin{array}{c}
e^{-i\theta/2}\\
-e^{i\theta/2}\\
-e^{i\theta/2}\\
e^{-i\theta/2}
\end{array}\right),\left|\chi_{--}\right\rangle =\frac{1}{2}\left(\begin{array}{c}
-e^{-i\theta/2}\\
e^{i\theta/2}\\
-e^{i\theta/2}\\
e^{-i\theta/2}
\end{array}\right),\label{eq:buracos int}
\end{equation}
 and, in the particle-hole representation we have
\begin{equation}
H_{ph}=\left(\begin{array}{cccc}
k & 0 & \Delta & 0\\
0 & -k & 0 & -\Delta\\
\Delta & 0 & k & 0\\
0 & -\Delta & 0 & -k
\end{array}\right).
\end{equation}
 The Hamiltonian now obeys $\mathcal{S}$ time reversal symmetry and
there is no energy gap ($\mathcal{S}$ is the sympletic time reversal symmetry
in the sense described in \cite{key-10}). However, there is another
interesting property that shows up. Because of the different translations
in energy of the two Dirac cones, there can be phenomena akin to the
Klein paradox in transmission problems. The particle-hole representation
reveals an effective interaction that favors the coupling of different
valley particles over the coupling of different valley holes states.\\

\subsubsection*{Spectrum in the presence of $\mathcal{V}_{I}$ and $\mathcal{V}_{z}$}

Two other Hamiltonians that possibly connect valleys
are given by (\ref{eq:v3}) and (\ref{eq:v4}) and therefore,
\begin{equation}
H=\left(\begin{array}{cccc}
0 & ke^{-i\theta} & \Delta & 0\\
ke^{i\theta} & 0 & 0 & \Delta\\
\Delta & 0 & 0 & ke^{i\theta}\\
0 & \Delta & ke^{-i\theta} & 0
\end{array}\right)
\end{equation}
 and
\begin{equation}
H=\left(\begin{array}{cccc}
0 & ke^{-i\theta} & \Delta & 0\\
ke^{i\theta} & 0 & 0 & -\Delta\\
\Delta & 0 & 0 & ke^{i\theta}\\
0 & -\Delta & ke^{-i\theta} & 0
\end{array}\right).
\end{equation}

These are of no actual interest to transmission problems. The dispersion
relations will be, respectively,
\begin{equation}
E_{\pm\pm x}=\pm\sqrt{k^{2}+\Delta^{2}\pm2k\Delta\cos\theta}
\end{equation}
 and
\begin{equation}
E_{\pm\pm y}=\pm\sqrt{k^{2}+\Delta^{2}\pm2k\Delta\sin\theta}.
\end{equation}
 These problems are related to each other by rotations of $\pi/2$
around the $z$ axis, and hence  do not introduce any new physics.
Besides, we see that the effect of this potential on the dispersion
relations of the particles reduces to just a change by $\pm\Delta$
in the $x$ (first case) or $y$ (second case) components of the particles'
momenta, which can be renormalized and the resulting quasi-particles behave no differently from free particles.
Therefore, in the regions affected by these potentials, the transmission
probabilities will be equal to one, as can be shown by solving the
semi-infinite or finite interacting regions problem. We will not deal
with these two cases any further and show, only for completeness,
the expressions for the spinors associated with the Hamiltonians above,
which respectively read:\begin{widetext}
\begin{equation}
\left|\chi_{-+}\right\rangle =\frac{1}{2}\left(\begin{array}{c}
1\\
-\frac{E_{++x}}{ke^{-i\theta}+\Delta}\\
-\frac{E_{++x}}{ke^{-i\theta}+\Delta}\\
1
\end{array}\right)\left|\chi_{++}\right\rangle =\frac{1}{2}\left(\begin{array}{c}
1\\
\frac{E_{++x}}{ke^{-i\theta}+\Delta}\\
\frac{E_{++x}}{ke^{-i\theta}+\Delta}\\
1
\end{array}\right)\left|\chi_{+-}\right\rangle =\frac{1}{2}\left(\begin{array}{c}
-1\\
-\frac{E_{+-x}}{ke^{-i\theta}-\Delta}\\
\frac{E_{+-x}}{ke^{-i\theta}-\Delta}\\
1
\end{array}\right)\left|\chi_{--}\right\rangle =\frac{1}{2}\left(\begin{array}{c}
-1\\
\frac{E_{+-x}}{ke^{-i\theta}-\Delta}\\
-\frac{E_{+-x}}{ke^{-i\theta}-\Delta}\\
1
\end{array}\right);
\end{equation}

\begin{equation}
\left|\chi_{-+}\right\rangle =\frac{1}{2}\left(\begin{array}{c}
1\\
-\frac{E_{++y}}{ke^{-i\theta}+\Delta}\\
-\frac{E_{++y}}{ke^{-i\theta}+\Delta}\\
1
\end{array}\right)\left|\chi_{++}\right\rangle =\frac{1}{2}\left(\begin{array}{c}
1\\
\frac{E_{++y}}{ke^{-i\theta}+\Delta}\\
\frac{E_{++y}}{ke^{-i\theta}+\Delta}\\
1
\end{array}\right)\left|\chi_{+-}\right\rangle =\frac{1}{2}\left(\begin{array}{c}
-1\\
-\frac{E_{+-y}}{ke^{-i\theta}-\Delta}\\
\frac{E_{+-y}}{ke^{-i\theta}-\Delta}\\
1
\end{array}\right)\left|\chi_{--}\right\rangle =\frac{1}{2}\left(\begin{array}{c}
-1\\
\frac{E_{+-y}}{ke^{-i\theta}-\Delta}\\
-\frac{E_{+-y}}{ke^{-i\theta}-\Delta}\\
1
\end{array}\right).
\end{equation}

\end{widetext}

\section{The Transmission Problem in 2D. "Barriers" and "Steps"}

From now onwards, we will be dealing with situations in which there
is transmission of electronic waves from "free" electron graphene
into semi-infinite or finite regions of "disturbed" graphene,
and also reflection back into the original region. We call these two
situations step and barrier problems, respectively.

Although there has been extensive use of wavefunction matching to
describe problems of electronic behavior in graphene (some examples
can be seen in \cite{key-11} and \cite{key-12}), we would like
to make some comments of our own. We shall start by briefly talking
about the simple 2D scattering from straight interfaces. Although
it is a simple problem it is very useful to establish the terminology
we employ in the more complex cases. All problems of transmission
begin with the determination of the wave functions in different regions
and the enforcement of the boundary conditions they obey. We assume
that the particle in medium I has positive energy, with momentum on
the Dirac cone $\mathbf{K}$ and moves to the right. Since both types
of interactions connect states of the different Dirac cones, conservation
of the valley "flavor" need not  take place any more and the
states corresponding to medium I must be given by \begin{widetext}
\begin{equation}
\Psi_{I}=\left[\frac{1}{\sqrt{2}}\left(\begin{array}{c}
e^{-i\theta/2}\\
e^{i\theta/2}\\
0\\
0
\end{array}\right)+\frac{r_{1}}{\sqrt{2}}\left(\begin{array}{c}
-ie^{i\theta/2}\\
ie^{-i\theta/2}\\
0\\
0
\end{array}\right)+\frac{r_{2}}{\sqrt{2}}\left(\begin{array}{c}
0\\
0\\
ie^{-i\theta/2}\\
-ie^{i\theta/2}
\end{array}\right)\right]e^{i\mathbf{k}\cdot\mathbf{r}}.
\end{equation}
 \end{widetext} Moreover, if the problem is of a barrier type, we
should be able to find in medium III the particles with momenta around
both valleys, independently of its flavor in medium I. Conservation
of energy demands that these particles in medium III must also have
positive energy and then we have
\begin{equation}
\Psi_{III}=\left[\frac{t_{1}}{\sqrt{2}}\left(\begin{array}{c}
e^{-i\theta/2}\\
e^{i\theta/2}\\
0\\
0
\end{array}\right)+\frac{t_{2}}{\sqrt{2}}\left(\begin{array}{c}
0\\
0\\
e^{i\theta/2}\\
e^{-i\theta/2}
\end{array}\right)\right]e^{i\mathbf{k}\cdot\mathbf{r}}.
\end{equation}
 The states accessible to the particles in medium II depend on the
dispersion relations of the Hamiltonian and we will deal with each
specific case in the next sections.

Since the Hamiltonians are of first order in $\hat{\mathbf{P}}$,
we need only to invoke the continuity of the wavefunctions at the
interfaces of the different media to be able to determine the coefficients
$r_{i}$ and $t_{i}$. However, some care must be taken when evaluating
the transmission and reflection amplitudes in the step problems. These
must be determined by the conservation of the probability current
in the direction normal to the interface (in our case, the $x$ direction).
Beginning with the continuity equation we have, for medium $i=\mbox{I},\,\mbox{II},\,\mbox{III}$,
\begin{equation}
\nabla\cdot\mathbf{J}_{i}=-\frac{\partial\rho_{i}}{\partial t}
\end{equation}
 which in the stationary regime implies that $\rho_{i}=\left|\Psi_{i}\right|^{2}$
is time independent and we have
\begin{equation}
\nabla\cdot\mathbf{J_{i}}=0.
\end{equation}
 Now, since our problem is translational invariant along the $y$
direction, ${\mathbf{J}}_{i}$ is independent of the $y$ variable and
it finally reduces to a one dimensional conservation problem
\begin{equation}
J_{ix}=const.
\end{equation}
 Although the conservation of the probability current is relevant
only in one dimension, the coefficients $r_{i}$ and $t_{i}$ and
the eigenstates (with which we will calculate the probability current)
all depend on the angles that the momenta of the particles make with
the normal to the interfaces in each medium. Since the states are
equal to each other at the interfaces (boundary conditions), we have
that
\begin{equation}
J_{Ix}=J_{IIx}=J_{IIIx},\label{eq:eqJ}
\end{equation}
 the last equality happening only for barrier problems.

We calculate the probability current in each medium by taking the
mean values of the current operator defined by
\begin{equation}
\mathbf{J}=-\frac{\delta H}{\delta\mathbf{A}}=e\vec{\sigma},
\end{equation}
 with the states $\Psi_{i}$, where $\mathbf{A}$ is the electromagnetic
vector potential. Notice that depending on the type of interaction
(for instance,  if it is momentum dependent), the expression for the current
will not assume this usual simple form and the definition through
the functional derivative must be used to find the correct expression.
The $x$ component of the probability currents in media I and III
will be always the same in our problems and are found to be
\begin{eqnarray}
J_{Ix} & = & \left[1-\left|r_{1}\right|^{2}-\left|r_{2}\right|^{2}\right]\cos\theta\label{eq:JIx}
\end{eqnarray}
 and
\begin{equation}
J_{IIIx}=\left[\left|t_{1}\right|^{2}+\left|t_{2}\right|^{2}\right]\cos\theta.
\end{equation}
 With these, we see that, for barrier problems
\begin{equation}
J_{Ix}=J_{IIIx}\Rightarrow1=\left|r_{1}\right|^{2}+\left|r_{2}\right|^{2}+\left|t_{1}\right|^{2}+\left|t_{2}\right|^{2},
\end{equation}
 and we recover the usual result. In step problems, we will have different
coefficients, depending on the states involved in the scattering at
the interface.

\section{Semi-infinite Regions: Step Problem}

\subsubsection*{$\mathcal{V}_{y}$ interaction}

This problem is very much alike the usual quantum mechanical problem
of transmission through a potential energy step smaller than the particle's
energy. In Fig. 3, we show the energetics of the problem and we easily
see that for $k<\Delta$, $k$ being the energy of the incident particle,
there will be no transmission, and hence we need the energy of the
incident particle to be larger than $\Delta$. Moreover, conservation
of energy and momentum along the $y$ direction demands that

\begin{figure}[tb]

\begin{centering}
\includegraphics[scale=0.7]{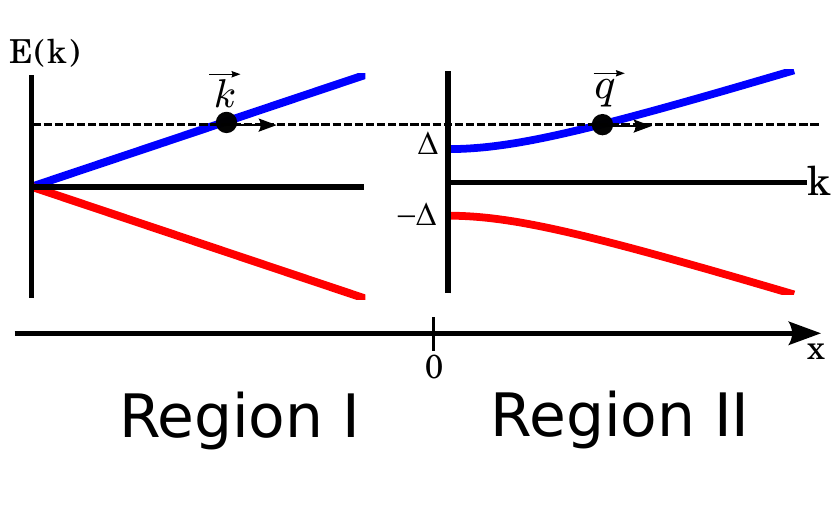}
\par\end{centering}

\caption{(color on-line) Schematics of the step transmission problem for $\mathcal{V}_{y}$
Hamiltonian. The red (light gray) curves are associated with hole
states and the blue (dark gray) ones with particles states.}
\end{figure}

\begin{eqnarray}
k & = & \sqrt{q^{2}+\Delta^{2}}\label{eq:consE1}\\
k\sin\theta & = & q\sin\alpha.\label{eq:snell}
\end{eqnarray}
 where $\alpha$ is the angle between the $y$ and $x$ components of the
wavevector of the transmitted wave. The transmission can take place
with any of the two degenerate states of positive energy, and, consequently,
the boundary conditions at the interface leads us to the system
\[
\frac{1}{\sqrt{2}}\left(\begin{array}{c}
e^{-i\theta/2}\\
e^{i\theta/2}\\
0\\
0
\end{array}\right)+\frac{r_{1}}{\sqrt{2}}\left(\begin{array}{c}
-ie^{i\theta/2}\\
ie^{-i\theta/2}\\
0\\
0
\end{array}\right)+\frac{r_{2}}{\sqrt{2}}\left(\begin{array}{c}
0\\
0\\
ie^{-i\theta/2}\\
-ie^{i\theta/2}
\end{array}\right)
\]

\begin{equation}
=\frac{t_{1}}{\sqrt{2}}\left(\begin{array}{c}
A_{+}e^{-i\alpha/2}\\
A_{+}e^{i\alpha/2}\\
A_{-}e^{i\alpha/2}\\
-A_{-}e^{-i\alpha/2}
\end{array}\right)+\frac{t_{2}}{\sqrt{2}}\left(\begin{array}{c}
A_{-}e^{-i\alpha/2}\\
-A_{-}e^{i\alpha/2}\\
-A_{+}e^{i\alpha/2}\\
-A_{+}e^{-i\alpha/2}
\end{array}\right).
\end{equation}
 Its solutions are
\begin{eqnarray}
r_{1} & = & \frac{q\sin\alpha-k\sin\theta}{k+q\cos(\alpha+\theta)};\label{eq:r1}\\
r_{2} & = & -i\Delta\frac{\cos\theta}{k+q\cos(\alpha+\theta)};\label{eq:r2}\\
t_{1} & = & \sqrt{2k(k+q)}\frac{\cos\theta\cos\left(\frac{\alpha+\theta}{2}\right)}{k+q\cos(\alpha+\theta)};\label{eq:t1}\\
t_{2} & = & i\sqrt{2k(k-q)}\frac{\cos\theta\sin\left(\frac{\alpha+\theta}{2}\right)}{k+q\cos(\alpha+\theta)};\label{eq:t2}
\end{eqnarray}
 which shows that there are reflection and transmission to all the
energy degenerate states. Similar expressions for the transmission
and reflection amplitudes and probabilities (which follow below) are
always found when dealing with wave-function matching in graphene
and have previously been derived in the literature \cite{key-11,key-12}.

The evaluation of the probability current in medium II reveals that
$J_{IIx}=\frac{q}{k}\left(\left|t_{1}\right|^{2}+\left|t_{2}\right|^{2}\right)\cos\alpha$.
Hence, by (\ref{eq:eqJ}) and (\ref{eq:JIx}),
\begin{equation}
T=\left(\left|t_{1}\right|^{2}+\left|t_{2}\right|^{2}\right)\frac{q}{k}\frac{\cos\alpha}{\cos\theta}.
\end{equation}
 \textcolor{black}{{} We can use \eqref{eq:consE1}, the coefficients
\eqref{eq:r1}-\eqref{eq:t2} and the definition of $\alpha=\tan^{-1}(q_{y}/q_{x})$
, to express this transmission probability (TP) as a function of the given parameters of the
disturbing potential and of the incident wave. }

\textcolor{black}{So we get
\begin{equation}
T=\frac{2\sqrt{1-\eta^{2}}\cos\alpha}{1+\sqrt{1-\eta^{2}}\cos\alpha},
\end{equation}
 with the angle of emergence given by}

\textcolor{black}{{}
\begin{equation}
\alpha=\tan^{-1}\left(\frac{\sin\theta}{\sqrt{\cos^{2}\theta-\eta^{2}}}\right),\label{eq:alpha}
\end{equation}
 where we defined $\eta\equiv\Delta/k$ (this procedure will be repeated
throughout the whole paper in a more succinct fashion, unless some special warning is necessary).}

One should be careful about some details of these solutions. First
of all, one notices from (\ref{eq:alpha}) that there are situations
in which, even if the energy of the particle is greater than the interaction
energy, for large enough incidence angles, $\alpha$ will be complex,
also leading to a complex TP, which has no physical meaning. This
phenomenon is analogous to total reflection in electromagnetism. This
happens because although in the transmission and reflection probabilities
(TP and RP) expressions the transmission and reflection amplitudes
appear only in squared moduli, the probability current also depends
on the angles. Making the substitution of imaginary $q_{x}$ and $\alpha$
first in the wavefunctions and then proceeding with the calculation
of $J_{IIx}$ reveals that actually $J_{IIx}=0$, as expected. This
is the same as happens for scattering by step potentials in usual
quantum mechanics when we deal with energies above and below the step
energy.

\begin{figure*}
\begin{centering}
\includegraphics[scale=0.7]{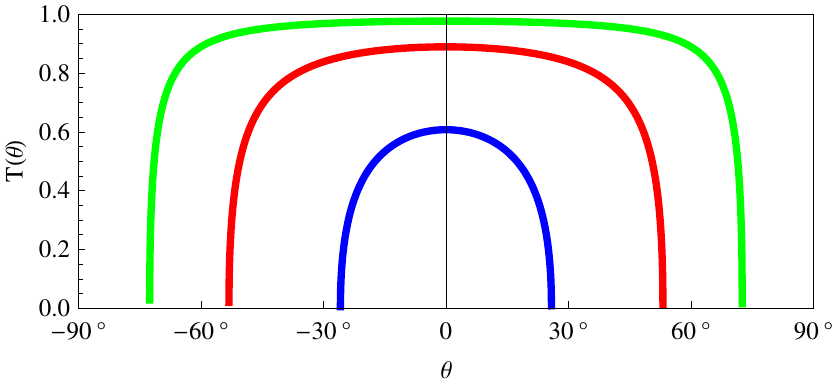} \includegraphics[scale=0.7]{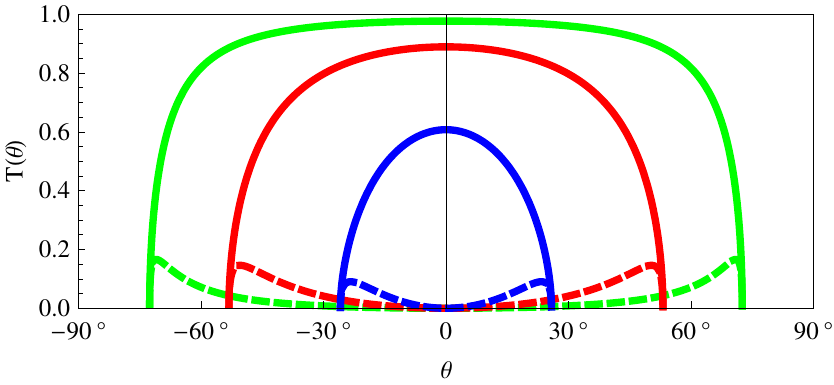}
\par\end{centering}

\begin{centering}
\includegraphics[scale=0.7]{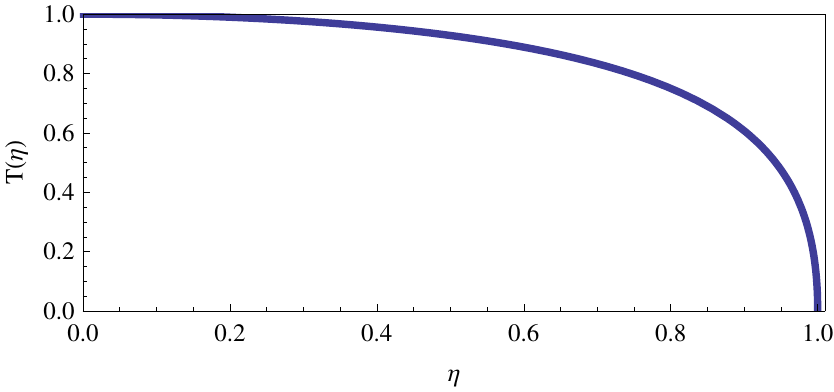} \includegraphics[scale=0.7]{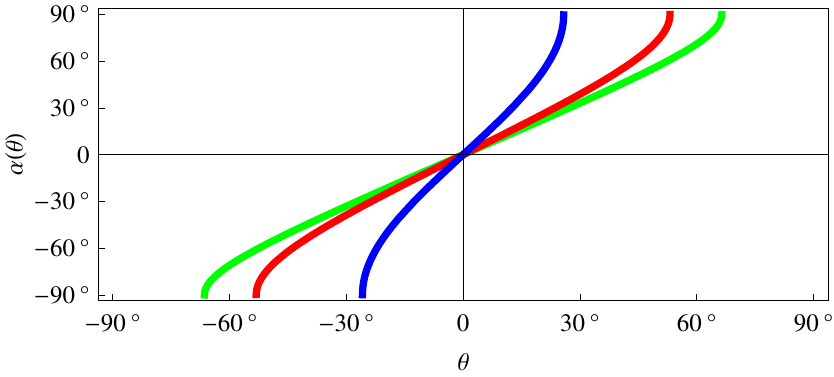}
\par\end{centering}

\caption{(color on-line) Behavior of the TP for the $\mathcal{V}_{y}$ step
problem. In the upper and lower figures on the right, we have $\eta$
equals to 0.3, 0.6 and 0.9,for the green (light gray), red (mid gray)
and blue (dark gray) curves, respectively. (Upper left) Total TP as
a function of the angle of incidence. (Upper right) We separate here
the TP associated with states 1 (full line) and 2(dashed line). (Lower
left) TP for normal incidence as a function of the ratio of the energies
of interaction and incidence. It's important to notice that the contributions
here come only from state 1. (Lower right) "Refraction angle"
as a function of the incident angle for the same values of $\eta$
above.}
\end{figure*}

In Fig. 4, we show the general behavior of the TP. Some interesting
features of the problem arise now, as we see that state 2 only contributes
to transmission about the extreme values of the allowed angles. As
we raise the barrier in relation to the particle's energy, we see
that the probability of transmission drops to zero, as expected. We
also note that there is a focalization of the beam, with respect to
the possible incidence angles which allow for transmission with non-zero
probability. Looking at the behavior of the "angles of refraction",
which characterize the direction of propagation of the wave-fronts
of the spatial parts of the wavefunction, as a function of the angle
of incidence, we see that for small values they are about the same,
at least for low $\eta$. One should notice that these angles need
not have actual relation to the direction of propagation of the probability
current, since the spinors also depend on the momenta and this affects
the mean value of the current operator (it is connected to the spinors
through the $\vec{\sigma}$ matrices).

\subsubsection*{$\mathcal{V}_{x}$ interaction}

The scheme of the dispersion relations for this physical situation
is shown in Fig. 5.

\begin{figure*}[t]

\begin{centering}
\includegraphics[scale=0.7]{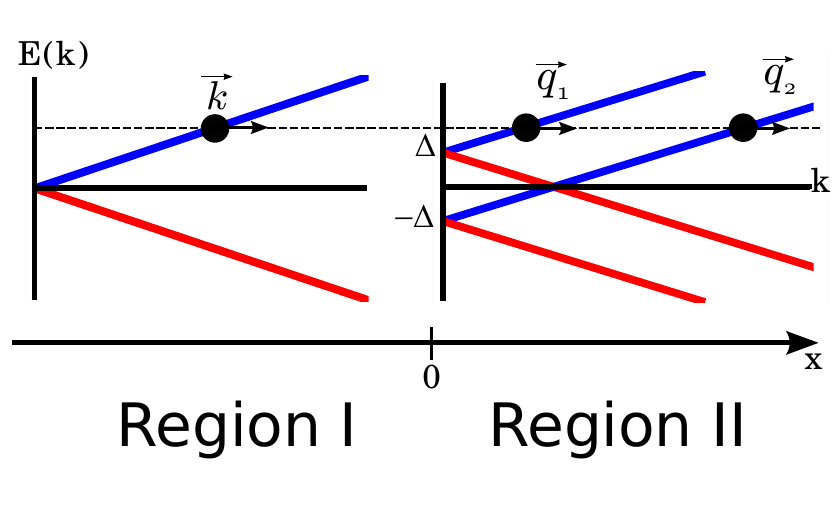} \includegraphics[scale=0.7]{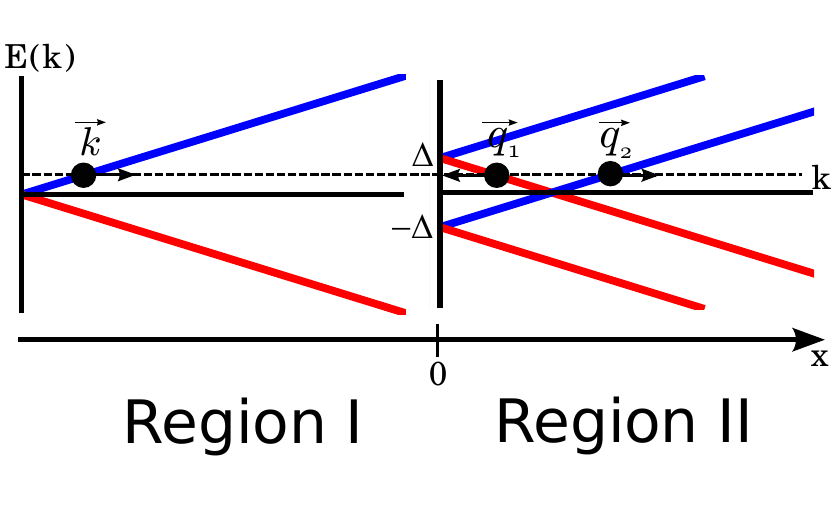}
\par\end{centering}

\caption{(color on-line) Scheme of the dispersion relation for the step transmission
problem for Hamiltonian $\mathcal{V}_{x}$. The red (light gray) curves
are associated with hole states and the blue (dark gray) ones with
particles states. (left) Transmission for $k>\Delta$. (right) Transmission
for $k<\Delta$. }
\end{figure*}

We notice that there are two non-degenerate bands accessible to the
incident particles (we will call these the states 1 and 2), which
lead to particles moving with different momenta inside the medium,
due to the conservation of energy. This will lead to two different
angles $\alpha_{i}$, $i=1,\,2$, for the emergent particles on the
right side of the interface, related to the angle of incidence $\theta$
by
\begin{equation}
q_{i}\sin\alpha_{i}=k\sin\theta.\label{eq:snelli}
\end{equation}
 There can also be two different cases, which will change the states
accessible to the emergent particles, namely $k>\Delta$ and $k<\Delta$.
These considerations lead to the following situations:

\paragraph*{i) $k>\Delta$}

In this case (Fig. 5 left), the accessible states in medium II are
given by $\left|\chi_{++}\right\rangle $ and $\left|\chi_{+-}\right\rangle $
from equation \eqref{eq:partics int} and we have
\begin{equation}
k=q_{1}+\Delta=q_{2}-\Delta\label{eq:consE2}
\end{equation}
 and the boundary condition at the interface leads to\begin{widetext}

\begin{equation}
\frac{1}{\sqrt{2}}\left(\begin{array}{c}
e^{-i\theta/2}\\
e^{i\theta/2}\\
0\\
0
\end{array}\right)+\frac{r_{1}}{\sqrt{2}}\left(\begin{array}{c}
-ie^{i\theta/2}\\
ie^{-i\theta/2}\\
0\\
0
\end{array}\right)+\frac{r_{2}}{\sqrt{2}}\left(\begin{array}{c}
0\\
0\\
ie^{-i\theta/2}\\
-ie^{i\theta/2}
\end{array}\right)=\frac{t_{1}}{2}\left(\begin{array}{c}
e^{-i\alpha_{1}/2}\\
e^{i\alpha_{1}/2}\\
e^{i\alpha_{1}/2}\\
e^{-i\alpha_{1}/2}
\end{array}\right)+\frac{t_{2}}{2}\left(\begin{array}{c}
e^{-i\alpha_{2}/2}\\
e^{i\alpha_{2}/2}\\
-e^{i\alpha_{2}/2}\\
-e^{-i\alpha_{2}/2}
\end{array}\right).
\end{equation}
 \end{widetext}

The solution of the linear system in this case is
\begin{eqnarray}
r_{1} & = & \frac{1}{2}\frac{\left[\sin\left(\frac{\alpha_{1}+\alpha_{2}}{2}\right)-\cos\left(\frac{\alpha_{1}-\alpha_{2}}{2}\right)\sin\theta\right]}{\cos\left(\frac{\alpha_{1}+\theta}{2}\right)\cos\left(\frac{\alpha_{2}+\theta}{2}\right)},\\
r_{2} & = & \frac{1}{2}\frac{\sin\left(\frac{\alpha_{1}-\alpha_{2}}{2}\right)\cos\theta}{\cos\left(\frac{\alpha_{1}+\theta}{2}\right)\sin\left(\frac{\alpha_{2}+\theta}{2}\right)},\\
t_{1} & = & \frac{\cos\theta}{\sqrt{2}\cos\left(\frac{\alpha_{1}+\theta}{2}\right)},\\
t_{2} & = & \frac{\cos\theta}{\sqrt{2}\cos\left(\frac{\alpha_{2}+\theta}{2}\right)}.
\end{eqnarray}

The probability current in the $x$ direction is given by
\begin{equation}
J_{IIx}=e\left(\left|t_{1}\right|^{2}\cos\alpha_{1}+\left|t_{2}\right|^{2}\cos\alpha_{2}\right),
\end{equation}
 leading to the transmission coefficient
\begin{equation}
T=\frac{\cos\alpha_{1}}{\cos\theta}\left|t_{1}\right|^{2}+\frac{\cos\alpha_{2}}{\cos\theta}\left|t_{2}\right|^{2}.
\end{equation}
 Substituting the transmission amplitudes of the solution gives
\begin{equation}
T=\cos\theta\left\{ \frac{\cos\alpha_{1}}{1+\cos[\alpha_{1}+\theta]}+\frac{\cos\alpha_{2}}{1+\cos[\alpha_{2}+\theta]}\right\} .\label{eq:transmit}
\end{equation}
 Proceeding in the same way as in the last section, we get
\begin{eqnarray}
\alpha_{1} & = & \tan^{-1}\left(\frac{\sin\theta}{\sqrt{\left(1-\eta\right)^{2}-\sin^{2}\theta}}\right);\\
\alpha_{2} & = & \tan^{-1}\left(\frac{\sin\theta}{\sqrt{\left(1+\eta\right)^{2}-\sin^{2}\theta}}\right);
\end{eqnarray}
 with which, one can plot the behavior of the TP, analogously to what
we have shown in Fig. 6.

Conservation of the $y$ component of the particle's momentum would
reveal that, in analogy to the classical electromagnetic case , particles
of band 1 feel the medium II "less refractive" and, consequently,
are fully reflected. However, for particles of band 2, the medium
is "more refractive" and there will always be transmission.

We notice that the main contribution to the total TP of states of
band 1 (those subject to total reflection) comes mostly from small
angles $\theta$, so the current must be almost normal to the interface
if one expects to transmit these particles. Looking at the "angles
of refraction" as a function of the angle of incidence, one can
easily recognize which band is subject to total reflection.

\begin{figure*}[t]

\begin{centering}
\includegraphics[scale=0.7]{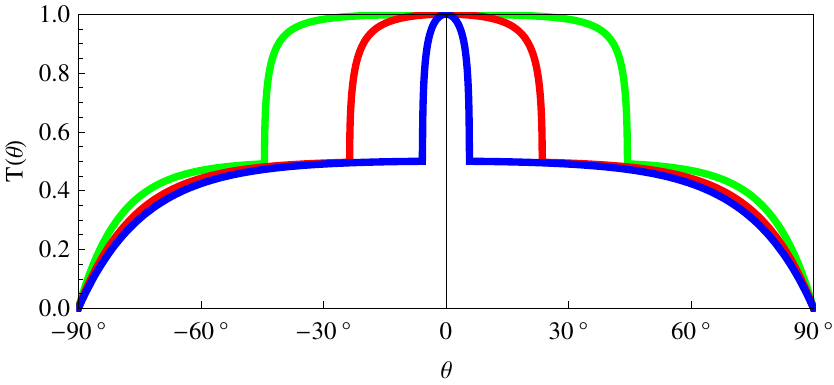}\includegraphics[scale=0.7]{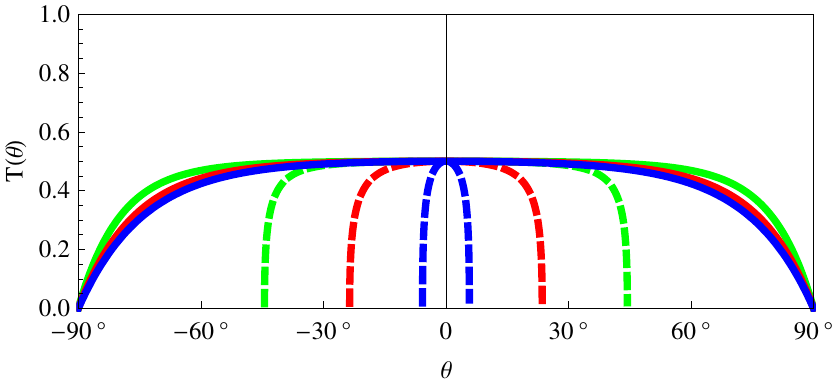}\includegraphics[scale=0.7]{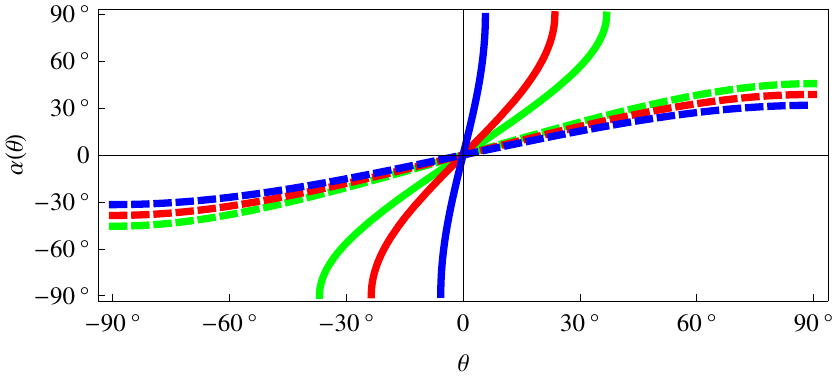}
\par\end{centering}

\caption{(color on-line) Behavior of the TP for the $\mathcal{V}_{x}$ $k>\Delta$
step problem. The curves plotted are for $\eta$ equals to 0.3, 0.6
and 0.9 corresponding to green (light gray), red (mid gray) and blue
(dark gray) respectively. (Left) Total TP in function of the angle
of incidence. (Middle) Contributions to the TP given by the separate
states 1 (full line) and 2(dashed line). (Right) "Refraction angles"
as a function of the incidence angle for the same values of $\eta$
above. Full lines correspond to $\alpha_{1}$ and dashed lines to
$\alpha_{2}$.}
\end{figure*}

\paragraph*{ ii) $k<\Delta$}

In this case, we have the possibility of conversion of a particle
state into a hole state, as shown in the right diagram of Fig. 5,
and the conservation of energy gives us
\begin{equation}
k=-q_{1}+\Delta=q_{2}-\Delta.
\end{equation}
 The linear system to be solved involves now the states $\left|\chi_{-+}\right\rangle $
and $\left|\chi_{+-}\right\rangle $, from \eqref{eq:partics int}
and \eqref{eq:buracos int} respectively, and reads\begin{widetext}
\begin{equation}
\frac{1}{\sqrt{2}}\left(\begin{array}{c}
e^{-i\theta/2}\\
e^{i\theta/2}\\
0\\
0
\end{array}\right)+\frac{r_{1}}{\sqrt{2}}\left(\begin{array}{c}
-ie^{i\theta/2}\\
ie^{-i\theta/2}\\
0\\
0
\end{array}\right)+\frac{r_{2}}{\sqrt{2}}\left(\begin{array}{c}
0\\
0\\
ie^{-i\theta/2}\\
-ie^{i\theta/2}
\end{array}\right)=\frac{t_{1}}{2}\left(\begin{array}{c}
e^{-i\alpha_{1}/2}\\
-e^{i\alpha_{1}/2}\\
-e^{i\alpha_{1}/2}\\
e^{-i\alpha_{1}/2}
\end{array}\right)+\frac{t_{2}}{2}\left(\begin{array}{c}
e^{-i\alpha_{2}/2}\\
e^{i\alpha_{2}/2}\\
-e^{i\alpha_{2}/2}\\
-e^{-i\alpha_{2}/2}
\end{array}\right).
\end{equation}
 \end{widetext}Its solution is
\begin{equation}
r_{1}=-\frac{1}{2}\frac{\left[\cos\left(\frac{\alpha_{1}+\alpha_{2}}{2}\right)+\sin\left(\frac{\alpha_{1}-\alpha_{2}}{2}\right)\sin\theta\right]}{\sin\left(\frac{\alpha_{1}+\theta}{2}\right)\cos\left(\frac{\alpha_{2}+\theta}{2}\right)}
\end{equation}

\begin{equation}
r_{2}=-\frac{1}{2}\frac{\cos\left(\frac{\alpha_{1}-\alpha_{2}}{2}\right)\cos\theta}{\sin\left(\frac{\alpha_{1}+\theta}{2}\right)\cos\left(\frac{\alpha_{2}+\theta}{2}\right)}
\end{equation}

\begin{equation}
t_{1}=i\frac{\cos\theta}{\sqrt{2}\sin\left(\frac{\alpha_{1}+\theta}{2}\right)}
\end{equation}

\begin{equation}
t_{2}=\frac{\cos\theta}{\sqrt{2}\sin\left(\frac{\alpha_{2}+\theta}{2}\right)},
\end{equation}
 and the probability current and TP are given by
\begin{eqnarray}
J_{IIx} & = & e\left(-\left|t_{1}\right|^{2}\cos\alpha_{1}+\left|t_{2}\right|^{2}\cos\alpha_{2}\right),\\
T & = & -\frac{\cos\alpha_{1}}{\cos\theta}\left|t_{1}\right|^{2}+\frac{\cos\alpha_{2}}{\cos\theta}\left|t_{2}\right|^{2},
\end{eqnarray}
 which, after proper substitution of the coefficients leads to
\begin{equation}
T=\cos\theta\left\{ -\frac{\cos\alpha_{1}}{1-\cos[\alpha_{1}+\theta]}+\frac{\cos\alpha_{2}}{1+\cos[\alpha_{2}+\theta]}\right\} .
\end{equation}
 One can see from this expression the expected appearance of the Klein
paradox, by looking at the denominator of this expression's first
term in the case of normal incidence. The solution of this is the
well-known argument that the hole-like particle must have an inverted
momentum, so that it will continue traveling from left to right in
the expected direction \cite{key-8}. For such, we demand
\begin{eqnarray}
\alpha_{1} & =\pi+ & \tan^{-1}\left(\frac{\sin\theta}{\sqrt{\left(1-\eta\right)^{2}-\sin^{2}\theta}}\right),\\
\alpha_{2} & = & \tan^{-1}\left(\frac{\sin\theta}{\sqrt{\left(1+\eta\right)^{2}-\sin^{2}\theta}}\right).
\end{eqnarray}
 We notice that, with these conventions for the angles, the expression
(\ref{eq:transmit}) is actually valid independently of the energy
of the particles and their interaction energy, but we will keep the
solutions separated. The general behavior of the TP for this case
is shown in Fig. 7.

We notice that as we keep raising the barrier (or lowering the particle's
energy), the contributions of both bands tend to become equal. It
is interesting to see that when $\eta=2$, particles of band 1 go
through the barrier as if there was nothing there, independently of
the angle of incidence. The behavior of the particle in band 1 as
a function of $\eta>1$ is such that the transmission starts around
zero for $\eta\approx1$ increases to uniform probability when $\eta=2$
and then tends to the same type of behavior of the particle in band
2.

From the behavior of the angles $\alpha_{1}$ and $\alpha_{2}$ as
a function of $\theta$ it is clear how, as we keep raising the barrier
(or lowering the energies), the direction of propagation of the wave-fronts
tend to become equal. For $\eta=2$ there is the explicit inversion
of behavior for the particle in band 1 and it stops being subject
to total reflection (notice that the rightmost plot of Fig. 7 is actually
of $\alpha_{1}-\pi$ but, since the holes move contrary to their momenta,
this is the actual direction of motion of the particle's wavefront).

\begin{figure*}[t]

\begin{centering}
\includegraphics[scale=0.7]{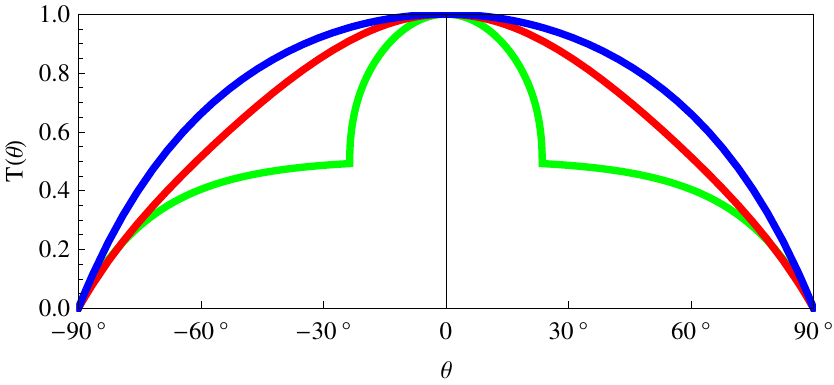}\includegraphics[scale=0.7]{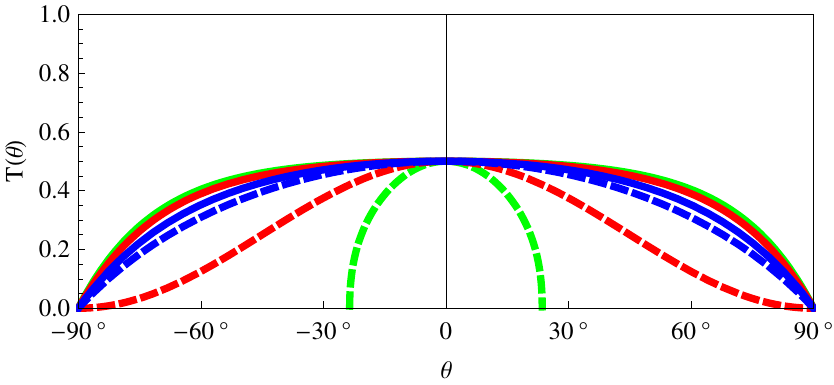}\includegraphics[scale=0.7]{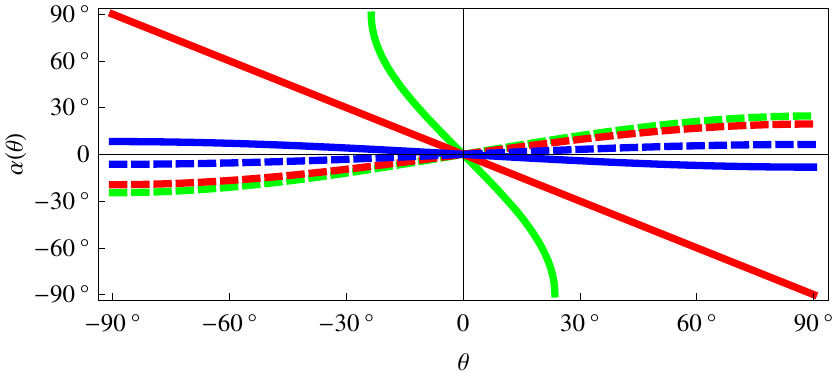}
\par\end{centering}

\caption{(color on-line) Behavior of the TP for the $\mathcal{V}_{x}$ $k<\Delta$
step problem. The curves plotted are for $\eta$ equals to 1.4, 2.0
and 8.0 corresponding to green (light gray),red (mid gray) and blue
(dark gray) respectively. (Left) Total TP in function of the angle
of incidence. (middle) Contributions to the TP given by the separate
states 1 (full lines) and 2(dashed lines). (right) behavior of $\alpha_{1}-\pi$(full
lines) and $\alpha_{2}$ (dashed lines) as a function of $\theta$.}
\end{figure*}

We can use the results of these last two cases to prepare a quasi-particle
current in the material in a definite quantum state or even in a known
linear combination of states. For the normal incidence problem, we
also see that there will be full transmission, independently of the
relation of the energy of the particle to the strength of the interaction.

\section{Finite Regions:Barrier Problems}

Now we revisit the problems of the last sections for the case of a
finite region of disturbed graphene. The mathematical developments
proceed almost in the same way as before, the only difference being
that we need to match the wavefunctions from medium I and medium II
at $x=0$ and from medium II and medium III at $x=d$, $d$ being
the width of the barrier, which will lead to a more complex linear
system. Hence, we have the following situations.

\subsubsection*{$\mathcal{V}_{y}$ interaction}

The matching of the wave function leads to the system\begin{widetext}

\[
\frac{1}{\sqrt{2}}\left(\begin{array}{c}
e^{-i\theta/2}\\
e^{i\theta/2}\\
0\\
0
\end{array}\right)+\frac{r_{1}}{\sqrt{2}}\left(\begin{array}{c}
-ie^{i\theta/2}\\
ie^{-i\theta/2}\\
0\\
0
\end{array}\right)+\frac{r_{2}}{\sqrt{2}}\left(\begin{array}{c}
0\\
0\\
ie^{-i\theta/2}\\
-ie^{i\theta/2}
\end{array}\right)=
\]

\begin{equation}
\frac{A_{1}}{\sqrt{2}}\left(\begin{array}{c}
A_{+}e^{-i\alpha/2}\\
A_{+}e^{i\alpha/2}\\
A_{-}e^{i\alpha/2}\\
-A_{-}e^{-i\alpha/2}
\end{array}\right)+\frac{A_{2}}{\sqrt{2}}\left(\begin{array}{c}
A_{-}e^{-i\alpha/2}\\
-A_{-}e^{i\alpha/2}\\
-A_{+}e^{i\alpha/2}\\
-A_{+}e^{-i\alpha/2}
\end{array}\right)+\frac{B_{1}}{\sqrt{2}}\left(\begin{array}{c}
-iA_{+}e^{i\alpha/2}\\
iA_{+}e^{-i\alpha/2}\\
iA_{-}e^{-i\alpha/2}\\
iA_{-}e^{i\alpha/2}
\end{array}\right)+\frac{B_{2}}{\sqrt{2}}\left(\begin{array}{c}
-iA_{-}e^{i\alpha/2}\\
-iA_{-}e^{-i\alpha/2}\\
-iA_{+}e^{-i\alpha/2}\\
iA_{+}e^{i\alpha/2}
\end{array}\right);
\end{equation}

\begin{eqnarray*}
\left[\frac{A_{1}}{\sqrt{2}}\left(\begin{array}{c}
A_{+}e^{-i\alpha/2}\\
A_{+}e^{i\alpha/2}\\
A_{-}e^{i\alpha/2}\\
-A_{-}e^{-i\alpha/2}
\end{array}\right)+\frac{A_{2}}{\sqrt{2}}\left(\begin{array}{c}
A_{-}e^{-i\alpha/2}\\
-A_{-}e^{i\alpha/2}\\
-A_{+}e^{i\alpha/2}\\
-A_{+}e^{-i\alpha/2}
\end{array}\right)\right]e^{iq_{x}d}+\left[\frac{B_{1}}{\sqrt{2}}\left(\begin{array}{c}
-iA_{+}e^{i\alpha/2}\\
iA_{+}e^{-i\alpha/2}\\
iA_{-}e^{-i\alpha/2}\\
iA_{-}e^{i\alpha/2}
\end{array}\right)+\frac{B_{2}}{\sqrt{2}}\left(\begin{array}{c}
-iA_{-}e^{i\alpha/2}\\
-iA_{-}e^{-i\alpha/2}\\
-iA_{+}e^{-i\alpha/2}\\
iA_{+}e^{i\alpha/2}
\end{array}\right)\right]e^{-iq_{x}d} & =
\end{eqnarray*}

\begin{equation}
\left[\frac{t_{1}}{\sqrt{2}}\left(\begin{array}{c}
e^{-i\theta/2}\\
e^{i\theta/2}\\
0\\
0
\end{array}\right)+\frac{t_{2}}{\sqrt{2}}\left(\begin{array}{c}
0\\
0\\
e^{i\theta/2}\\
e^{-i\theta/2}
\end{array}\right)\right]e^{ik_{x}d}.
\end{equation}
 \end{widetext}Notice that in medium II, we can now have both transmitted
and reflected waves, because of the presence of two interfaces. We
can solve this to find the coefficients $A_{i}$ and $B_{i}$ in terms
of $t_{i}$ and then solve a problem analogous to those of the last
sections. The dependence of $J_{IIx}$ on the position variables can
be easily seen to vanish, as one should expect, and the resulting
transmission probability can be directly obtained,
\begin{equation}
T=\left|t_{1}\right|^{2}+\left|t_{2}\right|^{2}.
\end{equation}
 The expression for the coefficients are cumbersome, but we show the
ones corresponding to the $t_{i}$'s and $r_{i}$'s. They are:
\begin{eqnarray}
r_{1} & = & i\frac{\sin\left(dq_{x}\right)(-q\sin\alpha+k\sin\theta)}{q\cos\left(dq_{x}\right)\cos\alpha\cos\theta+i\sin\left(dq_{x}\right)\left(k-q\sin\alpha\sin\theta\right)};\nonumber \\
r_{2} & = & -\frac{\sqrt{k^{2}-q^{2}}\cos\theta\sin\left(dq_{x}\right)}{q\cos\left(dq_{x}\right)\cos\alpha\cos\theta-i\sin\left(dq_{x}\right)\left(k-q\sin\alpha\sin\theta\right)};\nonumber \\
t_{1} & = & \frac{e^{-idk_{x}}q\cos\alpha\cos\theta}{q\cos\left(dq_{x}\right)\cos\alpha\cos\theta-i\sin\left(dq_{x}\right)\left(k-q\sin\alpha\sin\theta\right)};\nonumber \\
t_{2} & = & 0;
\end{eqnarray}

We are not able to write the expressions for the TPs in terms of the
ratio of the incident particle's energy to the potential strength
$\Delta$ anymore. In order to get an expression for the TPs, given
only in terms of known variables, we now use
\begin{eqnarray}
\alpha & = & \tan^{-1}\left(\frac{\sin\theta}{\sqrt{\cos^{2}\theta-\left(\frac{\Delta}{k}\right)^{2}}}\right);\\
q_{x} & = & k\sqrt{\cos^{2}\theta-\left(\frac{\Delta}{k}\right)^{2}},
\end{eqnarray}
finding
\begin{equation}
T=\frac{\left(k^{2}\cos^{2}\theta-\Delta^{2}\right)}{k^{2}\cos^{2}\theta-\Delta^{2}\cos^{2}\left(d\sqrt{k^{2}\cos^{2}\theta-\Delta^{2}}\right)}
\end{equation}
 The behavior of the TP is shown in Fig. 8. Since the emergent particles
come out in a medium equal to the one from which they enter, we do
not need to consider the incident and emergent angles, for they are
the same. We also have the situation of total reflection in this case
although, due to tunneling, the kinky behavior of the TP we had in
the step problems is not apparent any more.

One should notice  the oscillatory behavior generated by the self interference
of the wave reflected inside the medium II (Fabry-Perot like interferences).
In the middle plot, we realize that these oscillations are affected
mainly by the width of the barrier. We also notice that raising the
relative value of $\Delta$ over $k$ causes again a focalization
of the beam and gives rise to a resonant phenomenon of full TP.

These resonant points can be easily analyzed. The TP for $\theta=0$
reduces to
\begin{equation}
T|_{\theta=0}=\frac{k^{2}-\Delta^{2}}{k^{2}-\Delta^{2}\cos^{2}\left(dq\right)},\label{eq:tnormal}
\end{equation}
 whose maxima are seen to be given by
\begin{equation}
dq_{n}=n\pi\Rightarrow k_{n}=\sqrt{\left(\frac{n\pi}{d}\right)^{2}+\Delta^{2}},
\end{equation}
 which resembles the energy of massive relativistic particles in a
box. Expanding $t_{1}(\theta=0)$ for $q$ around $n\pi/d$, we also
get
\begin{equation}
t_{1}(\theta=0)\approx\pm i\frac{\Gamma_{n}/2}{k-k_{n}+i\Gamma_{n}/2},
\end{equation}
 where the $k_{n}'s$ are, as above, the energy levels of virtual
bound states whose inverse life-time (decay rates) are given by
\begin{equation}
\Gamma_{n}=\left.2\left(\frac{k}{q}\left(\frac{dq}{dk}\right)\right)\right|_{q=\frac{n\pi}{d}}=2\left(1+\left(\frac{\Delta d}{n\pi}\right)^{2}\right)^{-1}.
\end{equation}
 Similar life-times should appear for different values of the incident
angle. We see that if, for fixed $n$ and $\Delta$, the barrier width
gets larger, the particle life-time also increases. This is an expected
result since the mean number of times the particle should reflect
back and forth does not change and neither does its velocity, so the
time spent inside the barrier should be longer if it is larger. If
the interaction is stronger, i.e., the barrier is "higher", one
expect the particle's quantum behavior to be more important and therefore,
even if the particle's energy is above the barrier, it will spend
"more time" inside it and we expect a longer life-time.

\textcolor{black}{If in the normal incidence TP \eqref{eq:tnormal},
both the interaction $\Delta$ and the energy $k$ are small, we have,
up to second order in $k^{2}$ and $\Delta^{2}$, }

\textcolor{black}{{}
\begin{equation}
T|_{\theta=0}\approx\frac{k^{2}-\Delta^{2}}{k^{2}-\Delta^{2}+\Delta^{2}d^{2}\left(k^{2}-\Delta^{2}\right)}.
\end{equation}
 Decreasing the energy to arbitrarily low values leads to}

\textcolor{black}{{}
\begin{equation}
T|_{\theta=0}\approx\frac{1}{1+d^{2}\Delta^{2}}.
\end{equation}
 Hence, if the interaction energy is low enough (or even if the barrier
width is small enough, as can be easily checked), it is possible to
have transmission for arbitrarily low energies, even if smaller than
$\Delta$. This is a phenomenon analogous to tunneling in usual quantum
mechanics barrier problems. The wavefunction component dependent on
the $x$ direction becomes a real exponential, which can in some situations
penetrate in the other medium.}

\textcolor{black}{Notice that the transmission amplitude for states
2 $\left(t_{2}\right)$ is identically zero. This means that there
can be no flipping of the pseudo-spin due to the $\mathcal{V}_{y}$
coupling Hamiltonian. The situation is different for the $\mathcal{V}_{x}$
interaction, as will be shown in the next sections.}
\begin{figure*}[t]

\begin{centering}
\includegraphics[scale=0.7]{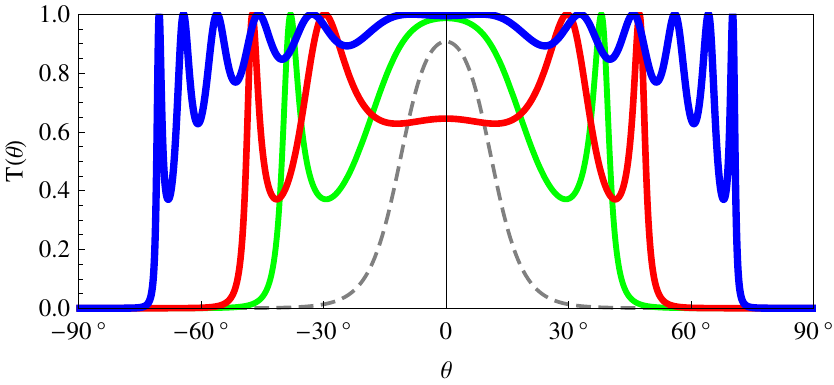}\includegraphics[scale=0.7]{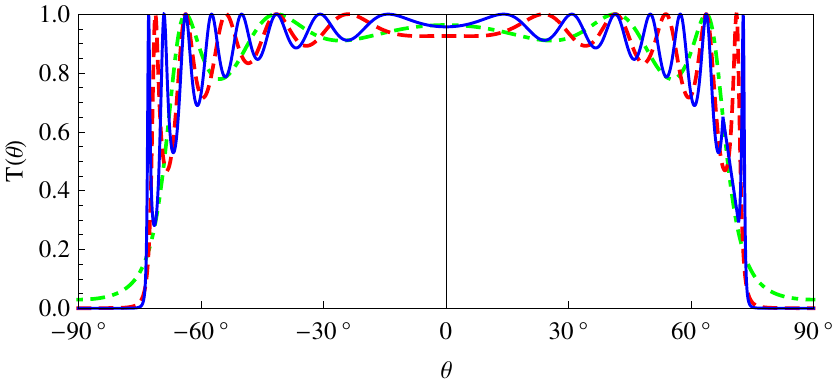}\includegraphics[scale=0.7]{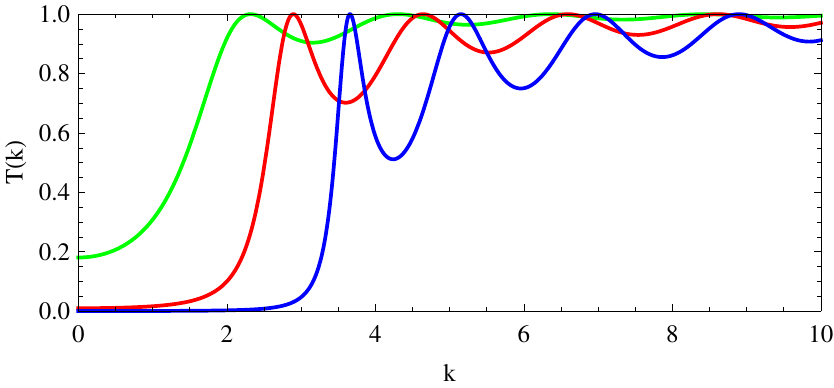}
\par\end{centering}

\caption{(color on-line) Behavior of the TP for the $\mathcal{V}_{y}$ barrier
problem. Since $t_{2}=0$, we only plot the total TP. On the left
we have $d=2.0$ and $\Delta=3.0$ fixed and vary the energy $k$
of the particles as $3.35$ (gray, dashed light gray), $4.3$ (green,
full line light gray), $5.0$ (red, mid gray) and $10.0$ (blue, dark
gray). In the middle, we fix $\Delta=4.45$ and $k=9.0$ and vary
the width $d$ through $1.0$(dashed green, light gray) , $2.0$(dashed
red, mid gray) and $3.0$(full line blue, dark gray). On the right,
we plot the normal TP as a function of $k$ for $d=1.5$ and $\Delta$
equals to $1.0$ (green, light gray), $2.0$ (red, mid gray), $3.0$
(blue, dark gray).}
\end{figure*}

\subsubsection*{$\mathcal{V}_{x}$ interaction}

\paragraph*{i)$k>\Delta$}

The linear system in this particular case is now \begin{widetext}
\[
\frac{1}{\sqrt{2}}\left(\begin{array}{c}
e^{-i\theta/2}\\
e^{i\theta/2}\\
0\\
0
\end{array}\right)+\frac{r_{1}}{\sqrt{2}}\left(\begin{array}{c}
-ie^{i\theta/2}\\
ie^{-i\theta/2}\\
0\\
0
\end{array}\right)+\frac{r_{2}}{\sqrt{2}}\left(\begin{array}{c}
0\\
0\\
ie^{-i\theta/2}\\
-ie^{i\theta/2}
\end{array}\right)=
\]

\begin{equation}
\frac{A_{1}}{2}\left(\begin{array}{c}
e^{-i\alpha_{1}/2}\\
e^{i\alpha_{1}/2}\\
e^{i\alpha_{1}/2}\\
e^{-i\alpha_{1}/2}
\end{array}\right)+\frac{B_{2}}{2}\left(\begin{array}{c}
e^{-i\alpha_{2}/2}\\
e^{i\alpha_{2}/2}\\
-e^{i\alpha_{2}/2}\\
-e^{-i\alpha_{2}/2}
\end{array}\right)+\frac{B_{1}}{2}\left(\begin{array}{c}
-ie^{i\alpha_{1}/2}\\
ie^{-i\alpha_{1}/2}\\
ie^{-i\alpha_{1}/2}\\
ie^{i\alpha_{1}/2}
\end{array}\right)+\frac{B_{2}}{2}\left(\begin{array}{c}
-ie^{i\alpha_{2}/2}\\
-ie^{-i\alpha_{2}/2}\\
-ie^{-i\alpha_{2}/2}\\
ie^{i\alpha_{2}/2}
\end{array}\right);
\end{equation}

\[
\frac{A_{1}}{2}\left(\begin{array}{c}
e^{-i\alpha_{1}/2}\\
e^{i\alpha_{1}/2}\\
e^{i\alpha_{1}/2}\\
e^{-i\alpha_{1}/2}
\end{array}\right)e^{iq_{1x}d}+\frac{B_{2}}{2}\left(\begin{array}{c}
e^{-i\alpha_{2}/2}\\
e^{i\alpha_{2}/2}\\
-e^{i\alpha_{2}/2}\\
-e^{-i\alpha_{2}/2}
\end{array}\right)e^{iq_{2x}d}+\frac{B_{1}}{2}\left(\begin{array}{c}
-ie^{i\alpha_{1}/2}\\
ie^{-i\alpha_{1}/2}\\
ie^{-i\alpha_{1}/2}\\
ie^{i\alpha_{1}/2}
\end{array}\right)e^{-iq_{1x}d}+\frac{B_{2}}{2}\left(\begin{array}{c}
-ie^{i\alpha_{2}/2}\\
-ie^{-i\alpha_{2}/2}\\
-ie^{-i\alpha_{2}/2}\\
ie^{i\alpha_{2}/2}
\end{array}\right)e^{-iq_{2x}d}=
\]

\begin{equation}
\left[\frac{t_{1}}{\sqrt{2}}\left(\begin{array}{c}
e^{-i\theta/2}\\
e^{i\theta/2}\\
0\\
0
\end{array}\right)+\frac{t_{2}}{\sqrt{2}}\left(\begin{array}{c}
0\\
0\\
e^{i\theta/2}\\
e^{-i\theta/2}
\end{array}\right)\right]e^{ik_{x}d}.
\end{equation}
 \end{widetext} The expressions for the solutions in this case are
even more cumbersome than in the previous case, although they still
resemble the expressions presented in the last section for this particular
situation. We will show only the different behavior of the TP patterns
related to them (Fig. 9). The angles $\alpha_{1}$ and $\alpha_{2}$
and the $x$ components of the momenta of states 1 and 2 are the same
of those of the corresponding case of the previous section. We see
that, for small angles, the resonant phenomena induced by the finite
barrier is irrelevant. There is some selection of states for larger
angles and only states of type 2 contribute to the central region.
For the normal incidence, we also have a perfectly resonant behavior
between the two TPs (which cause the total transmission be always
equal to 1), as a function of the width of the barrier, which can
be used to select the different states. There is no resonance for
fixed width and variable energies of incident particles.
\begin{figure*}[t]

\begin{centering}
\includegraphics[scale=0.7]{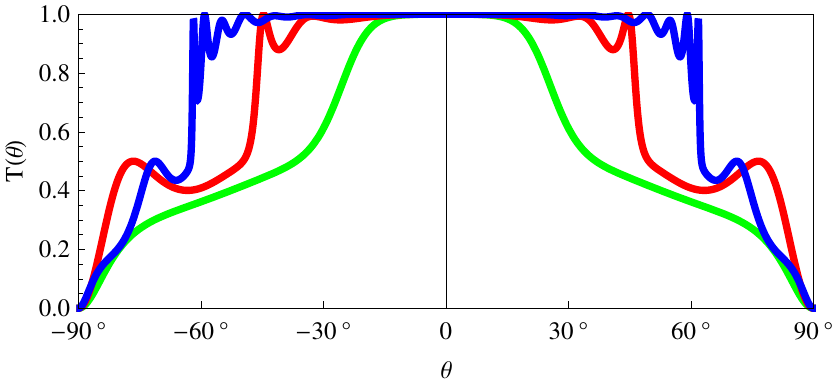} \includegraphics[scale=0.7]{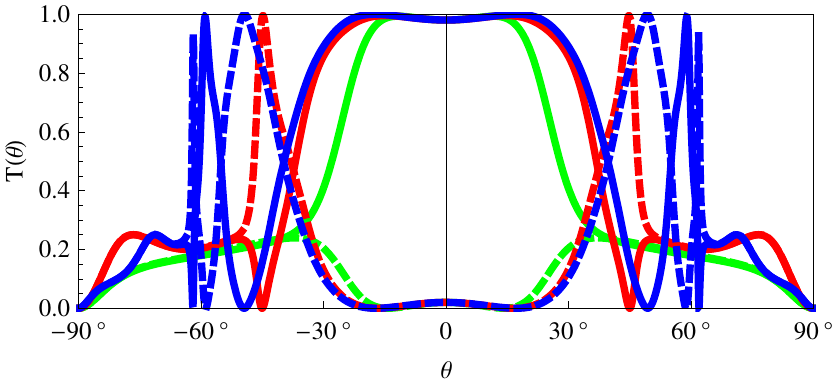}
\par\end{centering}

\begin{centering}
\includegraphics[scale=0.7]{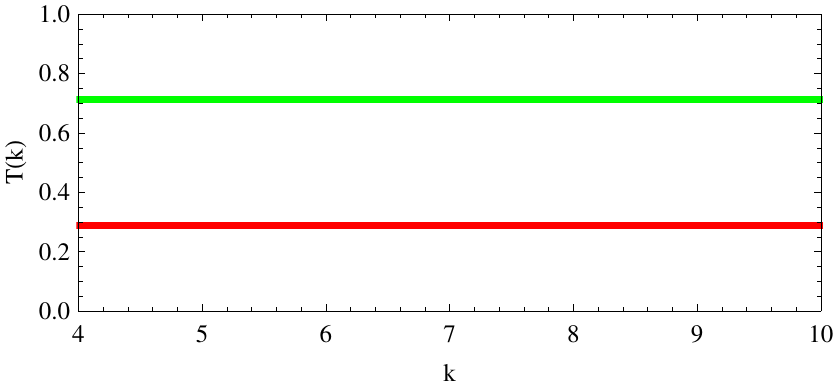} \includegraphics[scale=0.7]{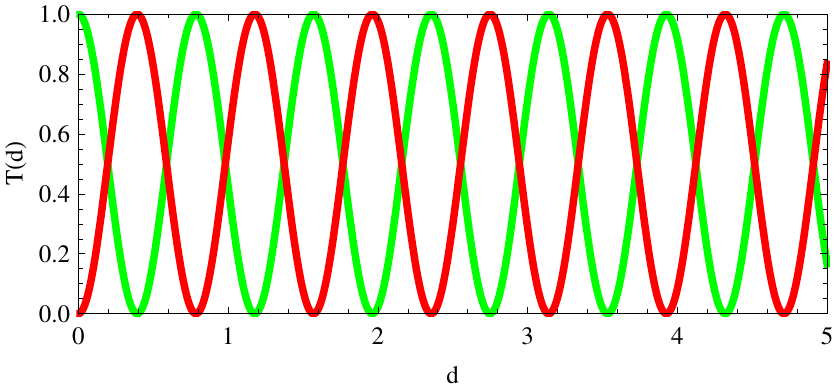}
\par\end{centering}

\caption{(color online) Behavior of the TP for the $\mathcal{V}_{x}$ $k>\Delta$
barrier problem. (upper left) We have $d=3.0$ and $\Delta=1.0$ fixed
and vary the energy $k$ of the particles as $2.0$ (green, light
gray), $4.0$ (red, mid gray) and $9.0$ (blue, dark gray). (upper
right) We separate here the contributions to the total TP from the
states 1(full lines) and 2 (dashed lines). (Lower left) Contributions
to the normal incidence given by state 1 (green, light gray) and state
2 (red, dark gray), for $\Delta=4.0$ and the same width of the barrier.
(Lower Right) Contributions to the normal incidence given by state
1 (green, light gray) and state 2 (red, dark gray), for $\Delta=4.0$
and $k=5$, for a varying width of the barrier.}
\end{figure*}

\paragraph*{ii) $k<\Delta$}

The system will change only through the transmitted state 1 within
region II and reads\begin{widetext}
\[
\frac{1}{\sqrt{2}}\left(\begin{array}{c}
e^{-i\theta/2}\\
e^{i\theta/2}\\
0\\
0
\end{array}\right)+\frac{r_{1}}{\sqrt{2}}\left(\begin{array}{c}
-ie^{i\theta/2}\\
ie^{-i\theta/2}\\
0\\
0
\end{array}\right)+\frac{r_{2}}{\sqrt{2}}\left(\begin{array}{c}
0\\
0\\
ie^{-i\theta/2}\\
-ie^{i\theta/2}
\end{array}\right)=
\]

\begin{equation}
\frac{A_{1}}{2}\left(\begin{array}{c}
e^{-i\alpha_{1}/2}\\
-e^{i\alpha_{1}/2}\\
-e^{i\alpha_{1}/2}\\
e^{-i\alpha_{1}/2}
\end{array}\right)+\frac{B_{2}}{2}\left(\begin{array}{c}
e^{-i\alpha_{2}/2}\\
e^{i\alpha_{2}/2}\\
-e^{i\alpha_{2}/2}\\
-e^{-i\alpha_{2}/2}
\end{array}\right)+\frac{B_{1}}{2}\left(\begin{array}{c}
-ie^{i\alpha_{1}/2}\\
-ie^{-i\alpha_{1}/2}\\
-ie^{-i\alpha_{1}/2}\\
-ie^{i\alpha_{1}/2}
\end{array}\right)+\frac{B_{2}}{2}\left(\begin{array}{c}
-ie^{i\alpha_{2}/2}\\
-ie^{-i\alpha_{2}/2}\\
-ie^{-i\alpha_{2}/2}\\
ie^{i\alpha_{2}/2}
\end{array}\right);
\end{equation}

\[
\frac{A_{1}}{2}\left(\begin{array}{c}
e^{-i\alpha_{1}/2}\\
-e^{i\alpha_{1}/2}\\
-e^{i\alpha_{1}/2}\\
e^{-i\alpha_{1}/2}
\end{array}\right)e^{iq_{1x}d}+\frac{B_{2}}{2}\left(\begin{array}{c}
e^{-i\alpha_{2}/2}\\
e^{i\alpha_{2}/2}\\
-e^{i\alpha_{2}/2}\\
-e^{-i\alpha_{2}/2}
\end{array}\right)e^{iq_{2x}d}+\frac{B_{1}}{2}\left(\begin{array}{c}
-ie^{i\alpha_{1}/2}\\
-ie^{-i\alpha_{1}/2}\\
-ie^{-i\alpha_{1}/2}\\
-ie^{i\alpha_{1}/2}
\end{array}\right)e^{-iq_{1x}d}+\frac{B_{2}}{2}\left(\begin{array}{c}
-ie^{i\alpha_{2}/2}\\
-ie^{-i\alpha_{2}/2}\\
-ie^{-i\alpha_{2}/2}\\
ie^{i\alpha_{2}/2}
\end{array}\right)e^{-iq_{2x}d}=
\]

\begin{equation}
\left[\frac{t_{1}}{\sqrt{2}}\left(\begin{array}{c}
e^{-i\theta/2}\\
e^{i\theta/2}\\
0\\
0
\end{array}\right)+\frac{t_{2}}{\sqrt{2}}\left(\begin{array}{c}
0\\
0\\
e^{i\theta/2}\\
e^{-i\theta/2}
\end{array}\right)\right]e^{ik_{x}d},
\end{equation}
 \end{widetext}whose solution gives the TP plotted in Fig. 10. Here
again we notice the already familiar behavior of selection of states
2 only in the central region. We notice that the solutions behave
just like a continuation of the $k>\Delta$ problem and we have no
energy resonances for the normal transmission problem. For small energy
we have a transmission probability with poorly defined peaks, although
if we look at the separate contributions of the two states, we see
that they are such that the maxima and minima are opposite, except
for normal transmission. Changing the size of the barrier, once again,
induces resonances.
\begin{figure*}[t]

\begin{centering}
\includegraphics[scale=0.7]{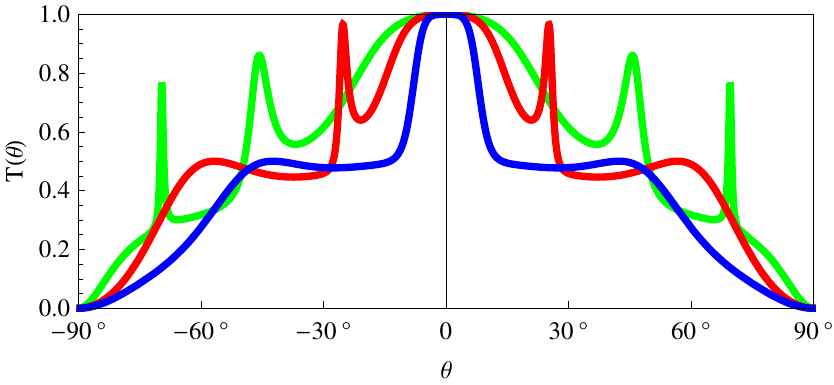} \includegraphics[scale=0.7]{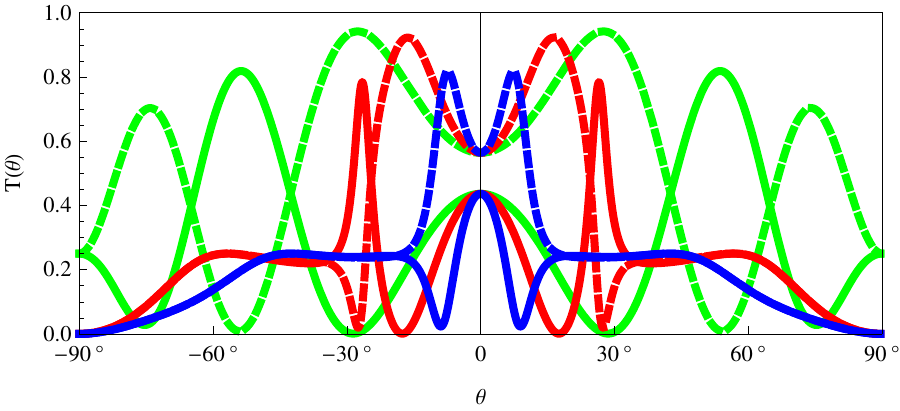}
\par\end{centering}

\begin{centering}
\includegraphics[scale=0.7]{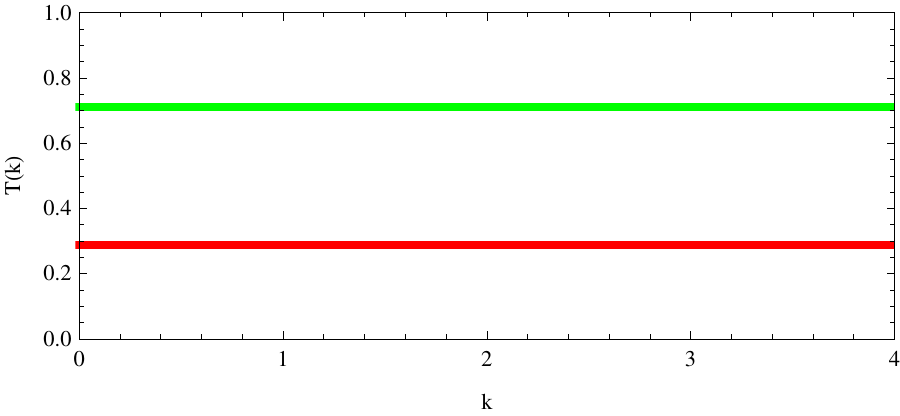} \includegraphics[scale=0.7]{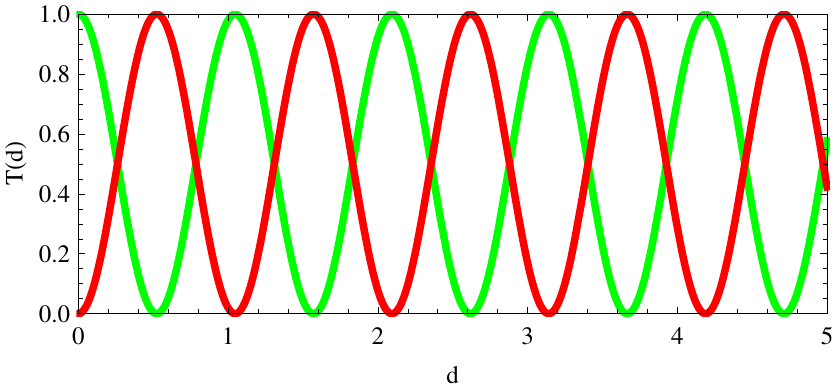}
\par\end{centering}

\caption{(color online) Behavior of the TP for the $\mathcal{V}_{x}$ $k<\Delta$
barrier problem . (Upper left) We have $d=3.0$ and $\Delta=6.0$
fixed and vary the energy $k$ of the particles as $3.0$ (green,
full line light gray), $4.0$ (red, mid gray) and $5.0$ (blue, dark
gray). In the middle, we separate the contributions to the total TP
from the states 1(full lines) and 2 (dashed lines) for the same values
of $d,\,\Delta,\, k$. (Lower left) The figure shows the contributions
to the normal incidence given by state 1 (green, light gray) and state
2 (red, dark gray), for $\Delta=4.0$. Notice the continuation of
this last graphic in comparison to the lower left graphic in figure
9. (Lower right) Contributions to the normal incidence given by state
1 (green, light gray) and state 2 (red, dark gray), for $\Delta=4.0$
and $k=3$ , for a varying width of the barrier.}
\end{figure*}

\section{Conductance and the Landauer Formalism}

We will try to use now the results obtained above to  access a measurable
quantity, namely, the conductance of the material due to the presence
of the barriers. We understand that our solution here might be an oversimplification,
since we are not dealing with disorder, finite size effects, contacts,
etc. Nevertheless, we believe that this might give us some insight on
the usefulness of results like ours. We shall use a reasoning based
on the Landauer formalism. The total transmitted current through our
sample should be given by
\begin{equation}
I\approx\left(\mu_{1}-\mu_{2}\right)\int dE\sum_{i}\mathcal{D}_{1}^{i}\left(E\right)J_{t}^{i}\left(E\right)\left[-\frac{\partial f_{0}\left(E\right)}{\partial E}\right],
\end{equation}
where $\mathcal{D}_{1}^{i}\left(E\right)$ is the one dimensional
density of states per unit length for channel $i$, $J_{t}^{i}\left(E\right)$
is the total transmitted current through our barrier through channel
$i$ and the sum is over all the channels. $\mu_{1}$ and $\mu_{2}$
are the respective chemical potentials of two particle reservoirs
thorough which the electrons will flow and $f_{0}$ is the Fermi distribution
with null chemical potential.

This leads to
\[
\frac{I}{\left(\mu_{1}-\mu_{2}\right)e}=\frac{I}{\Delta\phi}
\]
\begin{eqnarray}
G & \approx & e\int dE\sum_{i}\mathcal{D}_{1}^{i}\left(E\right)J_{t}^{i}\left(E\right)\left[-\frac{\partial f_{0}\left(E\right)}{\partial E}\right],
\end{eqnarray}
where $\Delta\phi$ is the potential difference and $G$ is the
conductance between the reservoirs.

The channels $i$ are given by the discrete values that $k_{y}$ would
assume in a finite sized sample, in such a way that the dispersion
relation is to be given by (putting back the $\hbar$'s)
\begin{eqnarray*}
E^{2} & = & p^{2}\Rightarrow E^{2}=p_{x}^{2}+\hbar^2\left(i\pi/L\right)^{2}\\
p_{x}^{2} & = & E^{2}-\hbar^2\left(i\pi/L\right)^{2}\\
p_{i}^{2} & \equiv & E^{2}-\epsilon_{i}^{2},
\end{eqnarray*}
where $\epsilon_{i}$ takes the value for $p_{y}$, $p_{i}$ the value
for $p_{x}$ and $i$ are integer numbers. Notice that $p_{i}$ would
give the energy for a Dirac particle moving in graphene in one dimension
at the channel $i$. The total current we are dealing with here, $J_{t}^{i}\left(E\right)$,
is normal to the barrier, the possibilities of different incident
angles is absorbed in the channels as we make clear below.

Therefore, with this prescription, the transmission probability for
the $\mathcal{V}_{y}$ interaction is given by
\begin{eqnarray}
T & = & \frac{\left(E^{2}\cos^{2}\theta-\Delta^{2}\right)}{E^{2}\cos^{2}\theta-\Delta^{2}\cos^{2}\left(d\sqrt{E^{2}\cos^{2}\theta-\Delta^{2}}\right)}\nonumber \\
 & = & \frac{\left(E^{2}-p_{y}^{2}-\Delta^{2}\right)}{E^{2}-p_{y}^{2}-\Delta^{2}\cos^{2}\left(d\sqrt{E^{2}-p_{y}^{2}-\Delta^{2}}\right)}\nonumber \\
\Rightarrow T^{i}\left(E\right) & = & \frac{\left(E^{2}-\epsilon_{i}^{2}-\Delta^{2}\right)}{E^{2}-\epsilon_{i}^{2}-\Delta^{2}\cos^{2}\left(d\sqrt{E^{2}-\epsilon_{i}^{2}-\Delta^{2}}\right)}\nonumber \\
\mbox{or }T\left(E\left(p_{i}\right)\right) & = & \frac{\left(p_{i}^{2}-\Delta^{2}\right)}{p_{i}^{2}-\Delta^{2}\cos^{2}\left(d\sqrt{p_{i}^{2}-\Delta^{2}}\right)},
\end{eqnarray}
where $E\left(p_{i}\right)=p_{i}$,which makes explicit our argument
that the transmissions are 1D-like, normal to the barrier, and separated
in different channels.

Now let's analyze the expression for the current.
We can write,
\begin{eqnarray}
\sum_{i}\mathcal{D}_{1}^{i}\left(E\right)J_{t}^{i}\left(E\right) & = & \sum_{i}J_{t}^{i}\left(E\right)\frac{1}{L}\sum_{p_{i}}\delta\left(E-E\left(p_{i}\right)\right)\nonumber \\
 & = & \frac{1}{L}\sum_{p_{i},i}\delta\left(E-E\left(p_{i}\right)\right)J_{t}^{i}\left(E\right),
\end{eqnarray}
and using the results of sec. IV for the transmission direction only,
$J_{t}^{i}\left(E\right)\equiv eT^{i}\left(E\right)\left|v_{x}\left(p_{i}\right)\right|=eT\left(E\left(p_{i}\right)\right)\left|v_{x}\left(p_{i}\right)\right|$
( the transmission is assumed to be along the $x$ direction and the
modulus is present to guarantee that only the positive $x$ direction
is to be considered), one has
\[
\sum_{i}\mathcal{D}_{1}^{i}\left(E\right)J_{t}^{i}\left(E\right)=
\]
\begin{eqnarray}
 & = & e\frac{1}{L}\sum_{p_{i},i}\delta\left(E-E\left(p_{i}\right)\right)T\left(E\left(p_{i}\right)\right)\left|v_{x}\left(p_{i}\right)\right|\nonumber \\
 & = & e\frac{1}{L}\sum_{p_{i},i}\delta\left(E-E\left(p_{i}\right)\right)T\left(E\right)\left|v_{x}\left(p_{i}\right)\right|\nonumber \\
 & = & e\frac{1}{L}T\left(E\right)\sum_{p_{i},i}\delta\left(E-E^{i}\left(p_{i}\right)\right)\left|v_{x}\left(p_{i}\right)\right|,
\end{eqnarray}
where $L$ is the length of the sample. Now we can stop looking at
the values of $p_{y}$ as different modes and sum over all the values
of the vector $\mathbf{k}$ ( in two dimensions) obtaining
\begin{eqnarray}
\sum_{i}\mathcal{D}_{1}^{i}\left(E\right)J_{t}^{i}\left(E\right) & = & e\frac{1}{L}T\left(E\right)\sum_{p_{i},i}\delta\left(E-E\left(p_{i}\right)\right)\left|v_{x}\left(p_{i}\right)\right|\nonumber \\
 & = & \frac{e}{\pi\hbar}T\left(E\right)\sum_{\mathbf{p}}\delta\left(E-E\left(\mathbf{k}\right)\right)\frac{\pi\hbar}{L}\left|v_{x}\left(\mathbf{p}\right)\right|\nonumber \\
 & = & \frac{e}{\pi\hbar}T\left(E\right)\underbrace{\sum_{\mathbf{p}}\delta\left(E-E\left(\mathbf{p}\right)\right)\frac{\pi\hbar}{L}\left|v_{x}\left(\mathbf{p}\right)\right|}_{\equiv M\left(E\right)}\nonumber \\
 & = & \frac{e}{\pi\hbar}T\left(E\right)M\left(E\right).
\end{eqnarray}
The function $M\left(E\right)$ defines our number of
transverse modes. It can be calculated by the  usual method, transforming
the sum in an integral and remembering to count the spin degeneracy.
Valley degeneracy should not be included because the electrons are chosen
in a very well defined valley, otherwise we should add another factor of
2 multiplying the final expression. It gives, for a sample of width $W$
\begin{equation}
M\left(E\right)=\frac{2W}{\pi\hbar}\left|E\right|.
\end{equation}
or, putting back the Fermi velocity(remember that in our units were such that $v_{F}=1$)
\begin{equation}
M\left(E\right)=\frac{2W}{\pi\hbar v_{F}}\left|E\right|.
\end{equation}

We can easily determine, now, the value of the conductance due to
our barrier. For zero temperature (in which case the integral is trivial) we
get,
\begin{equation}
G=\frac{e^{2}}{\pi\hbar}T\left(E_{F}\right)\frac{2W}{\pi\hbar v_{F}}E_{F}
\end{equation}
(notice that the units are correct).

Hence, we show that the conductance is expected to be proportional
to the TP in the normal direction with the energy set at $E_{F}$. As
we have shown and analyzed in the last section the behavior of this
TP we will neither bother writing the explicit expressions nor going through
the analysis of the graphic representation once again. The only comment we believe
it deserves is to notice that for the $\mathcal{V}_{x}$ case we expect
to be able to choose the contributions to the conductance for each
different cone, as a function of the width of the barrier. We also expect oscillatory
behavior of the conductance as we vary the energy in the case $\mathcal{V}_{y}$.

\section{Conclusions}

We analyzed the problem of electronic transmission in graphene through
interfaces between regions in which quasi-particles belonging to different
valleys (Dirac cones) interact or not. The relevant Hamiltonians we
have employed are seen to be able to either create a gap in the quasi-particle
spectrum or shift the Dirac cones with respect to each other. In the
latter case, hole states become available to positive energy incident
particles and the Klein paradox arises.

The behavior of the TP indicate that, for both Hamiltonians, in barrier
and step like problems, there is focalization of the incident beam.
In barrier problems, due to the behavior of the wavefunctions in each
different region, we see that it should be possible to have situations
in which we can enter the system with particles with momenta around
one of the Dirac cones and come out with a superposition of electronic
states about both cones. We could also act in the reverse way and
filter states from one specific cone out of a general superposition
of electronic states involving different valleys. If systems in which
the electron-electron interactions we proposed can be isolated, this
physical phenomenon could be useful for the development of "valleytronics",
without dealing with edge modes \cite{key-13,key-14}.

Another possibility is to use this kind of transmission to entangle
a pair of originally separable electronic states belonging to different
cones. This would be very useful if one wishes to employ a graphene
sheet in the development of quantum processors.

In order to observe these effects, one should measure the contact
resistance of small graphene samples placed on appropriate substrates
which would induce the desired electron-electron interaction. Some considerations
towards this were given in the last section, where we showed that the
behavior of the conductance, at very low temperatures, is expected to be
proportional to the TP normal to the given barrier.

Finally, a word of caution about the Hamiltonians we have used to
induce the coupling between different valleys. Although we have appealed
to general arguments to propose the phenomenological forms we have
employed, we do not yet know of any microscopic mechanism to deduce
them. Nevertheless, we believe that they can indeed be obtained from
a more microscopic approach and shall be investigating this possibility
in the near future.

\section*{Acknowledgments}

P. L. e S. Lopes would like to thank "Fundação de Amparo à Pesquisa
no Estado de São Paulo" (FAPESP, Brazil) for financial support under
Grant No. 2009/18336-0 and A.O.C. acknolwedges financial support from
FAPESP and CNPq through the Instituto Nacional de Ciência e Tecnologia
em Informação Quântica (INCT-IQ).

\end{document}